\documentclass[usenatbib,useAMS,usegraphicx]{mn2e}
\usepackage{times}
\usepackage[psamsfonts]{amssymb}
\usepackage{natbib}
\usepackage[latin1,utf8]{inputenc}
\usepackage{graphicx}
\usepackage{xspace}
\usepackage{dcolumn}
\usepackage{bm}
\usepackage{physics}

\newlength{\figwidth}
\figwidth=\columnwidth 

\newlength{\mw}
\newcommand{\matlab}{\textsc{Matlab}\textsuperscript{\textregistered}\xspace}

\newcommand{\VOISE}{\renewcommand{\VOISE}{VOISE\xspace}VOronoi Image SEgmentation (VOISE)\xspace}

\newcommand{\VR}{\renewcommand{\VR}{VR\xspace}Voronoi region (VR)\xspace}
\newcommand{\VD}{\renewcommand{\VD}{VD\xspace}Voronoi diagram (VD)\xspace}
\newcommand{\CVD}{\renewcommand{\CVD}{CVD\xspace}centroidal Voronoi diagram (CVD)\xspace}

\newcommand{\HST}{\renewcommand{\HST}{HST\xspace}Hubble Space Telescope
(HST)\xspace}
\newcommand{\STIS}{\renewcommand{\STIS}{STIS\xspace}Space Telescope imaging
spectrograph (STIS)\xspace}
\newcommand{\ACS}{\renewcommand{\ACS}{ACS\xspace}Advanced Camera for Surveys
(ACS)\xspace}
\newcommand{\SBC}{\renewcommand{\SBC}{SBC\xspace}Solar-Blind Channel
(SBC)\xspace}
\newcommand{\WFPC}{\renewcommand{\WFPC}{WFPC2\xspace}Wide-Field Planetary
Camera~2 (WFPC2)\xspace}

\newcommand{\IRTF}{\renewcommand{\IRTF}{IRTF\xspace}Infrared Telescope Facility (IRTF)\xspace}
\newcommand{\NSFCAM}{\renewcommand{\NSFCAM}{NSFCam\xspace}}
\newcommand{\NSFCAMTWO}{\renewcommand{\NSFCAMTWO}{NSFCam~2\xspace}}

\newcommand{\IRAF}{\renewcommand{\IRAF}{IRAF\xspace}Image Reduction and
Analysis Facility (IRAF)\xspace}

\newcommand{\UKIRT}{\renewcommand{\UKIRT}{UKIRT\xspace}United Kingdom Infra-Red Telescope (UKIRT)} 
\newcommand{\UIST}{\renewcommand{\UIST}{UIST\xspace}\UKIRT imager-spectrometer (UIST)\xspace}

\newcommand{\Cassini}{\textit{Cassini}\xspace}

\newcommand{\SPICE}{SPICE\xspace}

\newcommand{\UV}{\renewcommand{\UV}{UV\xspace}ultraviolet (UV)\xspace}
\newcommand{\IR}{\renewcommand{\IR}{IR\xspace}infrared (IR)\xspace}

\newcommand{\CML}{\renewcommand{\CML}{CML\xspace}Central Meridian Longitude
(CML)\xspace}

\renewcommand{\arcsec}{\ensuremath{\mathrm{arc\,sec}}}
\renewcommand{\arcmin}{\ensuremath{\mathrm{arc\,min}}}

\newcommand{\Eq}[1]{Eq.~(\ref{#1})}
\newcommand{\Eqs}[2]{Eqs.~(\ref{#1}--\ref{#2})}
\newcommand{\Fig}[1]{Fig.~\ref{#1}}
\newcommand{\Table}[1]{Table~{\ref{#1}}}
\newcommand{\Section}[1]{section~{\ref{#1}}}

\newcommand{\obs}{\ensuremath{\oplus}}

\newcommand{\latobs}{{\beta_\obs}}
\newcommand{\lonobs}{{\lambda_\obs}}
\newcommand{\dirobs}{{\vec{\delta}_\obs}}

\renewcommand{\sun}{\ensuremath{\odot}}

\newcommand{\latsun}{{\beta_\sun}}
\newcommand{\lonsun}{{\lambda_\sun}}
\newcommand{\dirsun}{{\vec{\delta}_\sun}}

\newcommand{\surfnormal}{{\hat{\vec{n}}_S}}
\newcommand{\limbnormal}{{\hat{\vec{n}}_L}}
\newcommand{\termnormal}{{\hat{\vec{n}}_T}}
\newcommand{\latlimb}{{\beta_{\limbnormal}}}
\newcommand{\latterm}{{\beta_{\termnormal}}}

\newcommand{\dlon}{\Delta\lambda}
\newcommand{\re}{r_e}
\newcommand{\rp}{r_p}
\newcommand{\xc}{x_c}
\newcommand{\yc}{y_c}

\newcommand{\ls}{\ensuremath{\mathcal{L}}}
\newcommand{\area}{\ensuremath{\mathcal{A}}}
\newcommand{\pt}[1]{\vec{#1}}

\newcommand{\comment}[1]{}

\title{A new method for determining geometry of planetary images}
\author[P. Guio and N. Achilleos]{P. Guio\thanks{E-mail:p.guio@ucl.ac.uk}
and N. Achilleos\\ Physics and Astronomy, University College London,
Gower Place, London, WC1E 6BT, United Kingdom}

\begin{document}

\date{\normalsize$ $Date: 2010/10/06 16:37:15 $ $,~ $ $Revision: 1.136 $ $}

\pagerange{\pageref{firstpage}--\pageref{lastpage}} \pubyear{2010}

\maketitle

\label{firstpage}

\begin{abstract}
This paper presents a novel semi-automatic image processing technique to
estimate accurately, and objectively, the disc parameters of a planetary body
on an astronomical image. The method relies on the detection of the limb and/or
the terminator of the planetary body with the \VOISE algorithm
\citep{guio:2009a}.
The resulting map of the segmentation is then used to
identify the visible boundary of the planetary disc. The segments comprising
this boundary are then used to perform a ``best'' fit to an algebraic
expression for the limb and/or terminator of the body. 
We find that we are able to locate the centre of the planetary disc with an
accuracy of a few tens of one pixel. The method thus represents a useful
processing stage for auroral ``imaging'' based studies.
\end{abstract}


\section{Introduction}

During the last two decades, the \HST has provided resolved images of both
Jupiter and Saturn in the \UV spectral region. Such images capture with
high sensitivity and high resolution, the spectacular auroral phenomena
occurring in the polar regions of the gas giants as a result of energetic
magnetospheric particles raining down onto the planet's upper atmosphere. 
Auroral images have become a particularly useful diagnostic tool for
morphological characterisations of the aurora and its boundaries.
This is a crucial prerequisite for identifying the aurora's physical origin
\citep[e.g.][]{prange:1996,prange:1998,grodent:2003a,grodent:2003b,clarke:2005,badman:2008,lamy:2009}.

Imaging is also complementary to in situ measurements of the plasma
environment provided, e.g.\  by the \Cassini spacecraft, currently orbiting
Saturn. Combining remote imaging with in situ data allows the study of 
magnetospheric processes and how they affect the
planet's upper atmosphere, and ionosphere via the planet's magnetic field
\citep{dougherty:1998,clarke:2002,bunce:2008,talboys:2009}, and
the footprint auroral emission of satellites
\citep{clarke:2002,bonfond:2007,wannawichian:2008}. Such studies require
accurate projection of the
geographic and geomagnetic coordinate systems of the planet onto the plane of
the two-dimensional image.
Auroral dynamics  can be studied using time series of images. For these 
purposes,
the location of the planet centre needs to be known accurately.
Unfortunately, \HST pointing parameters are not generally known with sufficient
accuracy for such applications. The precision of the guide star catalogue
together with the uncertainty in the start time of the tracking motion is on
the order of \unit[1]{\arcsec} while it is desired to have an accuracy of the
order of \unit[1]{pixel}, i.e.\ \unit[0.02\mbox{--}0.03]{\arcsec} for the
\STIS and \ACS instruments, in order to to locate any structure accurately or
to build polar projections of the auroral emissions.

In addition, ground-based observations with telescopes such
as the NASA \IRTF and \UKIRT, both located at Mauna Kea, Hawaii, have provided
images of Jupiter and Saturn with resolved auroral structures in the \IR
waveband. 
\IR images and spectra also allow the study of the dynamics and morphology
of the \chem{H_3^{+}} molecular ion, a principal component of giant
ionospheres \citep{miller:2006}.
Again, the location of the planet centre needs
to be known accurately to make use of these images, but for similar
reasons as the \HST case, the
pointing parameters are not known with sufficient accuracy for the images
from \IRTF and \UKIRT telescopes. The resolution 
of the \IRTF \NSFCAMTWO imaging facility and  the \UIST are
respectively of the order of \unit[0.04]{\arcsec\;pixel^{-1}} and 
\unit[0.12]{\arcsec\;pixel^{-1}} or better.

The problem of the location of the planet on auroral images has been
addressed by various authors and studies 
\citep[e.g.][]{bonfond:2007,nichols:2008,bonfond:2009} but to our knowledge no
published work provides any detailed description of the method used.  Here we
propose a novel semi-automatic method to estimate accurately and objectively
the position, size and orientation of a planetary body. The method consists
of three phases: (i)
detection of the limb of the planet disc using our image segmentation
algorithm \VOISE (see \citet{guio:2009a} for details), (ii) selection of points (Voronoi seeds) from the \VOISE
map that surround the limb, and (iii) nonlinear fit
(Levenberg-Marquardt algorithm) of the selected set of data from \VOISE to
a disc model. Phase (i) is performed once while phases (ii) and (iii) can be
repeated in order to improve the accuracy.

In \Section{sec:limbterm}, we give analytic expressions for the
projection of the limb and the terminator in the sky-plane. In
\Section{sec:method}, the method is developed. In
\Section{sec:applications} we illustrate our method on \IR images of
Jupiter collected with the \IRTF and \UKIRT telescopes.
We discuss the performance of our method and summarise our conclusions in
\Section{sec:discussion}.

\section{Limb and terminator equations}
\label{sec:limbterm}

A planet's pressure surface can be modelled by an ellipsoid, more precisely an oblate spheroid,
with equatorial radius (semi-major axis) $\re$ and polar radius
(semi-minor axis) $\rp$, where $\rp^2=\re^2(1{-}e^2)$, and $e$ is the
eccentricity of the spheroid. The parameters are readily available,
for instance, from the NASA Navigation and Ancillary Information Facility
\SPICE system \citep{acton:1996}. 
\begin{figure}
\includegraphics[width=0.99\figwidth]{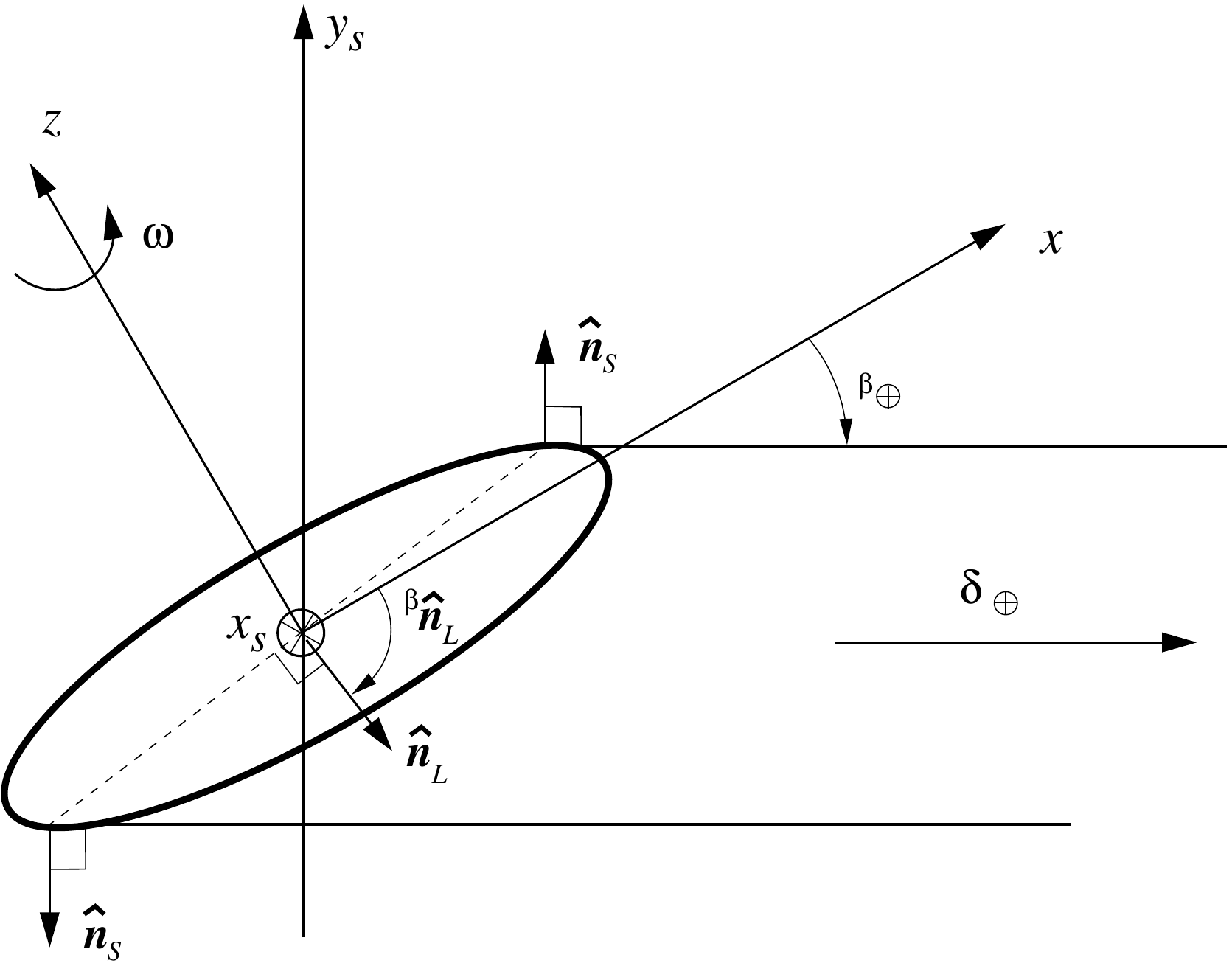}
\caption{Sketch of the geometry of the planet and the observer. The
eccentricity of the planetary ellipsoid is exaggerated for clarity.}
\label{fig:geom}
\end{figure}
The planet rotation vector is assumed to be along the $z$-axis with positive
angular velocity $\vec{\omega}$ as shown in \Fig{fig:geom}. 
Without loss of generality, we can further assume that the observer is
located in the $(x,z)$-plane, i.e.\ setting the longitude of the observer
$\lonobs{=}0$, and the observing direction as the vector
$\dirobs=(\cos\latobs,0,\sin\latobs)$ where $\latobs$ is the planetocentric 
latitude (sub-Earth latitude), the (negative) angle between the $x$-axis
and $\dirobs$ in \Fig{fig:geom}.
The limb of the planet consists of the points on the planet surface where
the local normal $\surfnormal$ is perpendicular to the observing direction
$\dirobs$, i.e.\ $\surfnormal\cdot\dirobs=0$.
The entire limb is contained in a single plane and is an ellipse, and
the normal vector to the plane containing the limb $\limbnormal$ has
coordinates $(\cos\latlimb,0,\sin\latlimb)$ where
\begin{align}
\cos\latlimb & = \frac{(1{-}e^2)\cos\latobs}
{\sqrt{(1{-}e^2)^2\cos^2\latobs+\sin^2\latobs}}\\
\sin\latlimb &= \frac{\sin\latobs}
{\sqrt{(1{-}e^2)^2\cos^2\latobs+\sin^2\latobs}}
\end{align}
The ellipse of the limb can then be projected onto the plane of the
sky $(x_s,y_s)$, i.e.\ on a plane perpendicular to the observing direction
$\dirobs$ (see \Fig{fig:geom}). The projection of the limb in the sky-plane
is also an ellipse 
with the following algebraic equation in the sky plane coordinate system
\begin{align}
\frac{x_s^2}{\re^2} +\frac{y_s^2}{\re^2(1{-}e^2\cos^2\latobs)} & = 1.
\label{eq:projlimb}
\end{align}
where the $y_s$-axis is chosen to lie in the $(x,z)$-plane.
It can be seen from \Eq{eq:projlimb} that the limb always appears with a
semi-major axis equal to the equatorial radius
of the planet and a semi-minor axis between the polar radius of the planet
(in the case of an equatorial view $\latobs=0$) and the equatorial radius of
the planet (in the case of a polar view $\latobs=\pm\pi/2$). 
Equivalently the eccentricity of the ellipse formed by the limb in the
sky is $e_L=e\cos\latobs$.

The terminator (boundary between day and night side) can be visualised as
the limb for the direction
corresponding to the location of the Sun $\dirsun$ (i.e.\ such that the
local normal $\termnormal$ is perpendicular to $\dirsun$) but projected onto
the sky plane along the observing direction $\dirobs$, i.e.\ directed
towards Earth. The Sun direction is defined by its planetocentric latitude 
$\latsun$ (sub-solar latitude) and its relative longitude (solar phase
angle) to the observing direction $\dlon =\lonsun{-}\lonobs$.
In the coordinate system $(x^\prime,y^\prime,z^\prime)$ where $z^\prime$ is
also aligned to the rotation axis of the planet but the plane
$(x^\prime,z^\prime)$ has been rotated about the planet rotation axis to
contain the Sun direction, the vector $\termnormal$ has coordinates
$(\cos\latterm,0,\sin\latterm)$ where
\begin{align}
\cos\latterm & = \frac{(1{-}e^2)\cos\latsun}
{\sqrt{(1{-}e^2)^2\cos^2\latsun+\sin^2\latsun}},\\
\sin\latterm & = \frac{\sin\latsun}
{\sqrt{(1{-}e^2)^2\cos^2\latsun+\sin^2\latsun}}
\end{align}
In the situation
where $\dirobs\times\termnormal=\vec{0}$, the limb and the terminator are
coincident. Otherwise the vector $\dirobs\times\termnormal$ is contained
in the sky plane (since it is perpendicular to $\dirobs$) and in the plane
of the terminator.
Therefore the projections onto the sky-plane of the limb and the terminator 
intersect at two points called the cusps, and the line joining
the two cusps has direction $\dirobs\times\termnormal$.

The two cusp points define the major axis of the ellipse formed by the
sky projection of the terminator. This shape is a tilted ellipse with
tilt angle $\theta_T$ with respect to the $x_s$-axis given by
\begin{multline}
\theta_T = \\
 \tan^{-1}\left(\frac{(1{-}e^2)\cos\latsun\sin\dlon}
{(1{-}e^2)\cos\latsun\cos\dlon\sin\latobs{-}\sin\latsun\cos\latobs}\right).
\end{multline}
The semi-major and semi-minor axes of the projection of the full terminator
(i.e.\ its visible and invisible parts) onto the sky-plane are
\begin{align}
a_T^2 &= (u^2{+}v^2)t_1^2 \\
b_T^2 &= (u^2{+}v^2)t_2^2.
\end{align}
where the vector $\vec{u}=(u,v)$ lies in the direction defined by
$\dirobs{\times}\termnormal$, 
\begin{align}
u & = (1{-}e^2)\cos\latsun\cos\dlon\sin\latobs-\sin\latsun\cos\latobs,\\
v &= (1{-}e^2)\cos\latsun\sin\dlon, 
\end{align}
and $t_1$ and $t_2$ are scalars analytically derivable from $u$ and $v$
\begin{align}
t_1^2 &= \frac{\re^2(1{-}e^2\cos^2\latobs)}{u^2(1{-}e^2\cos^2\latobs)+v^2}, \\
t_2^2 &= \frac{\re^2(1{-}e_\latsun^2)(ad{-}bc)^2}
{(dv{+}bu)^2(1{-}e_\latsun^2)+(cv{+}au)^2}
\end{align}
where $e_\latsun$ is the eccentricity of the ellipse formed by the terminator
in its own plane and can be expressed as function of the Sun planetocentric 
latitude $\latsun$ and the eccentricity of the spheroid $e$
\begin{align}
e_\latsun^2 & = e^2\left(1-\frac{\sin^2\latsun}{1-e^2\cos^2\latsun}\right),
\end{align}
and where
\begin{align}
a & = \cos\dlon, \\
b & = -\sin\latterm\sin\dlon, \\
c & = \sin\dlon\sin\latobs, \\
d & = \sin\latterm\cos\dlon\sin\latobs{+}\cos\latterm\cos\latobs
\end{align}

Finally the signed distance (as measured along $\termnormal$) between a point
$(x_T,y_T)$ of the projection of the terminator onto the
sky-plane, and the plane of the terminator itself is given by
\begin{multline}
D_T = (\cos\latterm\sin\dlon)x_T+\\
({-}\cos\latterm\cos\dlon\sin\latobs{+}\sin\latterm\cos\latobs)y_T,
\end{multline}
and points with $D_T{>}0$ belong to the visible terminator (from the
Earth observer's point of view) while points with
$D_T{<}0$ are hidden, and the case $D_T{=}0$ corresponds to cusp points.
Similarly the signed distance (measured along $\dirsun$) between a point
of the limb's projection onto the sky-plane $(x_L,y_L)$, and the plane
perpendicular to the direction of the Sun $\dirsun$ is given by
\begin{multline}
D_L = (\cos\dirsun\sin\dlon)x_L+\\
\left({-}\cos\dirsun\cos\dlon\sin\latobs{+}
\sin\dirsun\frac{\cos\latobs}{1{-}e^2}\right)y_L,
\end{multline}
and points such that $D_L{>}0$ belong to the illuminated limb while points
such that $D_L{<}0$ are in the shade, and the case $D_L{=}0$ corresponds
to cusp points.

\section{Planetary disc extraction method}
\label{sec:method}

As pointed out the proposed method to extract the orientation and shape of
the planetary disc from an image consists of three phases, described in the
following sections.

\subsection{Phase (i) \VOISE image reduction}

The first stage consists of partitioning the image into regions, i.e.\
simplify and/or change the representation of an image into something that is
more physically meaningful and easier to analyse.
\VOISE is a dynamic algorithm for partitioning the underlying pixel grid of
an image into regions according to a prescribed homogeneity criterion
\citep{guio:2009a}.
A \VOISE segmentation returns a map of the image in the form of a \VD where
each \VR is a polygon, within which the data are homogeneous with respect to prescribed criteria. When running the
\VOISE segmentation algorithm on an image of a planetary object we expect that
the transition region between the illuminated planet and the sky (i.e.\ the
limb or the terminator) consists of a ring of relatively small Voronoi polygons, 
indicating that at this region, the intensity is changing very quickly over
small spatial scales.
In this representation we can classify the ``ring'' of tiny Voronoi polygons
surrounding the larger central polygons as a cluster in itself that can be
used for fitting a terminator and/or a limb.
The map generated at the end of the \VOISE division phase provides the
largest number of seeds and smallest Voronoi polygons (see \citet{guio:2009a}
for more detail).

\subsection{Phase (ii) points selection}

The selection of seeds from the computed \VOISE map requires ``crude'' estimates
for the planet centre $(\xc,\yc)$,
the equatorial radius (semi-major axis) $\re$, the polar radius
(semi-minor axis) $\rp$ and the tilt angle $\alpha$.
The seeds from the segmentation are considered part of the
neighbourhood of the limb and/or terminator if they lie
inside a prescribed elliptic torus. A point belongs to the 
torus if its coordinates $(x,y)$ fulfil the following inequalities
\begin{align}
\varepsilon_m^2 
\leq\frac{{x^\prime}^2}{a^2}+\frac{{y^\prime}^2}{b^2}\leq\varepsilon_M^2,
\label{eq:select1}
\end{align}
where $\varepsilon_m$ and $\varepsilon_M$ (with 
$\varepsilon_m<1<\varepsilon_M$) represent the inner and outer
ellipses of the torus, and
where $(x^\prime,y^\prime)$ are obtained from 
$(x,y)$ by translation with ${-}(\xc,\yc)$ followed
by rotation with ${-}\alpha$, i.e.\
\begin{align}
\begin{bmatrix}x^\prime\\ y^\prime\end{bmatrix} = 
\begin{bmatrix}\cos\alpha & \sin\alpha\\ {-}\sin\alpha & \cos\alpha\end{bmatrix}
\begin{bmatrix}x{-}\xc\\ y{-}\yc\end{bmatrix}.
\label{eq:select2}
\end{align}
As well as satisfying the condition given by \Eqs{eq:select1}{eq:select2}, 
the polygons used
in the fitting procedure must also have a surface area smaller than a
prescribed value (equivalently a ``length scale'' $\ls$ smaller
than a maximum prescribed value $\ls_M$). The maximum length
scale $\ls_M$ has to be larger than the minimum distance between
seeds $d_m$ that is set for the \VOISE division phase.

It is also possible to filter out ``bands'' of polar angle in order to
remove those regions with auroral emission clearly outside the planetary limb
as illustrated in \Section{sec:applications}.

\subsection{Phase (iii) fitting of points}

Fitting of quadrics (such as circles and ellipses) to a given set of
points in the plane is a problem that arises in many application areas, e.g.\ 
computer graphics, pattern recognition, coordinate meteorology. Many algorithms 
minimise a quantity in some least-square sense. 
Such fitting algorithms for quadrics can be separated into categories of
``best fit'' (``geometric fit'') and ``algebraic fit'' 
\citep{gander:1994,fitzgibbon:1999}. In addition the clustering technique
is another technique to fit an ellipse, such as methods based on the 
Hough transform \citep{yuen:1989}.

In the ``best fit'', the quantity to minimise is the geometric distance
between the fitted curve and the
given set of points. In this case, curves may be represented in parametric form,
which is well suited for minimising the sum of the squares of the distances.

In the ``algebraic fit'' the curve is represented
algebraically, i.e.\ in the plane by an equation of the form $F(x,y)=0$.
If a point is on the curve, then its coordinates $(x,y)$ are a zero of the
function $F$ and represent an algebraic distance. These methods are usually
equivalent to solving a linear system of equations
subject to some constraint on the quadric coefficients
\citep{bookstein:1979,taubin:1991,fitzgibbon:1999}.
The constraint may be such that the optimal solution is computed directly,
and no iterations are needed.
The disadvantage of the ``algebraic fit'' is that we are uncertain what we
are minimising in a geometrical sense and in many cases those constraints
lead to fits which are not invariant under Euclidean transformations such as
translations and rotations, i.e.\ different coordinate systems produce
different fitted curves. Another limitation is that these methods can fit only
one primitive (or shape) at a time, therefore the data should be segmented
into a set of basic shapes before fitting each shape independently.
Nonetheless the algebraic solution is useful as an initial guess for
the geometrical fit.

We have developed a ``best fit'' tool based on the the Levenberg-Marquardt
method \citep{marquardt:1963}, for solving nonlinear least-square problems.
This method performs a minimisation
of the sum of the squares of the weighted distances between the $m$
selected seeds $\pt{s}_i$ ($i{=}1\ldots m$) from the Voronoi map, 
and the ``best''-fitting curve
with parametric representation $(x,y)=\vec{f}(\phi;\vec{p})$
\citep{bard:1974,gander:1994}. The minimisation consists in
adjusting iteratively the set of curve parameters $\{\phi_i\}$ 
(that locate the ``best'' points on the curve), together with 
the vector of global parameters $\vec{p}$ (that describe the global shape of
the curve). Mathematically the function to minimise is written
\begin{align}
Q(\phi_1,\phi_2,\ldots,\phi_m,\vec{p}) = \sum_{i=1}^m
\frac{\left\|\pt{s}_i-\vec{f}(\phi_i;\vec{p})\right\|^2}{\sigma_i^2}.
\label{eq:minfun}
\end{align}
The parametric representation $\vec{f}$ for a circle and an ellipse are given 
by \Eq{eq:circle} and \Eq{eq:ellipse} respectively. 
$\sigma_i^2$ represents the variance of the location of a given seed. 
An estimate
for this uncertainty can be readily computed as the mean distance from the
seed to all the points within the \VR.
Alternatively a length scale $\ls_i$ of the polygon can be inferred from
the square root of its surface area \citep{guio:2009a}, and can be thought
of as the ``average'' section length through the polygon in all directions. 
Thus considering the disc with same surface area $\area$ as the
polygon, the variance in distance of the disc points from its centre
is given by $\sigma^2=\area/(2\pi)=\ls^2/2$ (where surface area is used as
the weighting factor for variance). This expression can thus be used as
a reasonable approximation for the variance of the location of the seeds.

The Levenberg-Marquardt method is optimised to switch continuously from a
method which quickly approaches the minimum (the steepest descent method),
when far from the minimum, to a more precise but slower method (the Newton
method), when approaching the minimum. 

Finally we note that a priori knowledge about any of the parameters
can be used to constrain the fitting, otherwise an appropriate number of free
parameters may be simultaneously determined from the fitting procedure.

\subsubsection{Fitting a circle}

The parametric form used for the circle is given by
\begin{align}
\vec{f}(\phi;[\xc,\yc,r]) = 
\begin{bmatrix}\xc\\ \yc\end{bmatrix}{+}r
\begin{bmatrix}\cos\phi\\ \sin\phi\end{bmatrix}.
\label{eq:circle}
\end{align}
where $\xc$, $\yc$ and $r$ are respectively the coordinates of the centre and
the radius of the circle.
Note that the values of $\phi$ for each data point $\pt{s}_i$ are updated
along with the global parameters $\vec{p}=[\xc,\yc,r]$ at each iteration.
The parameter $\phi$ represents the polar angle, measured from the
$x$-axis, of the line joining the centre $(\xc,\yc)$ of the circle to the
point on the circle which is associated with the relevant data point.

\subsubsection{Fitting an ellipse}

The parametric form used for the ellipse is given by
\begin{multline}
\vec{f}(\phi;[\xc,\yc,a,b,\alpha]) = \\
\begin{bmatrix}\xc\\ \yc\end{bmatrix}{+}
\begin{bmatrix}\cos\alpha & {-}\sin\alpha\\ \sin\alpha &
\cos\alpha\end{bmatrix}
\begin{bmatrix}a\cos\phi\\ b\sin\phi\end{bmatrix}
\label{eq:ellipse}
\end{multline}
where $\xc$, $\yc$, $a$, $b$ and $\alpha$ are respectively the coordinates of
the centre, the semi-major axis, semi-minor axis and the tilt angle of
the ellipse (angle measured from the $x$-axis to the semi-major axis). 
In this case the parameter $\phi$ does not represent the polar
angle, measured from the $x$-axis, of the line joining the centre
$(\xc,\yc)$ of the ellipse to the point on the ellipse. The parameter
$\phi$ is sometimes referred to as eccentric anomaly and is related to
the polar angle $\theta$ by the following equation:
\begin{align}
b\tan\phi  = a\tan(\theta{-}\alpha).
\end{align}

\subsubsection{Fitting the limb and the terminator}

Note that the the limb and terminator can be represented in a single parametric
form $[x_{LT}(\phi),y_{LT}(\phi)]=\vec{f}_{LT}(\phi)$ using the equations
given in  \Section{sec:limbterm}.
This shape can be fitted by considering the three following transformations:
homothetic transformation with scale factor $c$, rotation with angle $\alpha$
and translation by $(\xc,\yc)$
\begin{multline}
\vec{f}(\phi;[\xc,\yc,c,\alpha]) = \\
\begin{bmatrix}\xc\\ \yc\end{bmatrix}{+}
c\begin{bmatrix}\cos\alpha & {-}\sin\alpha\\ \sin\alpha &
\cos\alpha\end{bmatrix}
\begin{bmatrix}x_{LT}(\phi)\\ y_{LT}(\phi)\end{bmatrix}
\label{eq:lt}
\end{multline}
In this case the global parameters for $\vec{f}_{LT}(\phi)$ are the
geometric parameters $\latobs$, $\latsun$, $\dlon$, $\re$, $\rp$
introduced in \Section{sec:limbterm} (as well
as the distance from the observer to the planet to convert projected length
into pixels units).  These parameters  may be
determined by e.g.\ \SPICE. The unknown global
parameters to optimise for $\vec{f}(\phi)$ are the planet centre $(\xc,\yc)$,
the scale factor $c$ and tilt angle $\alpha$.

\subsection{Algorithm}

Phase (i) is performed once while phases (ii) and (iii) can be iterated.
The iteration process improves the selection of seeds for the fit and
removes any ``outliers'' that might be included using the crude estimates
for the parameters, therefore improving the accuracy of the fit.
The tolerance on fractional improvement of $Q$ defined in \Eq{eq:minfun} is
set to $10^{-3}$. Such tolerance ensures convergence of the
Levenberg-Marquardt algorithm in a few iterations leading to an
accuracy of the estimate of the centre coordinates and the radii of the
order of one pixel or better. 

\section{Application to planetary images}
\label{sec:applications}

\subsection{Ellipse for complete planet}

\begin{figure}
\includegraphics[width=0.99\figwidth]{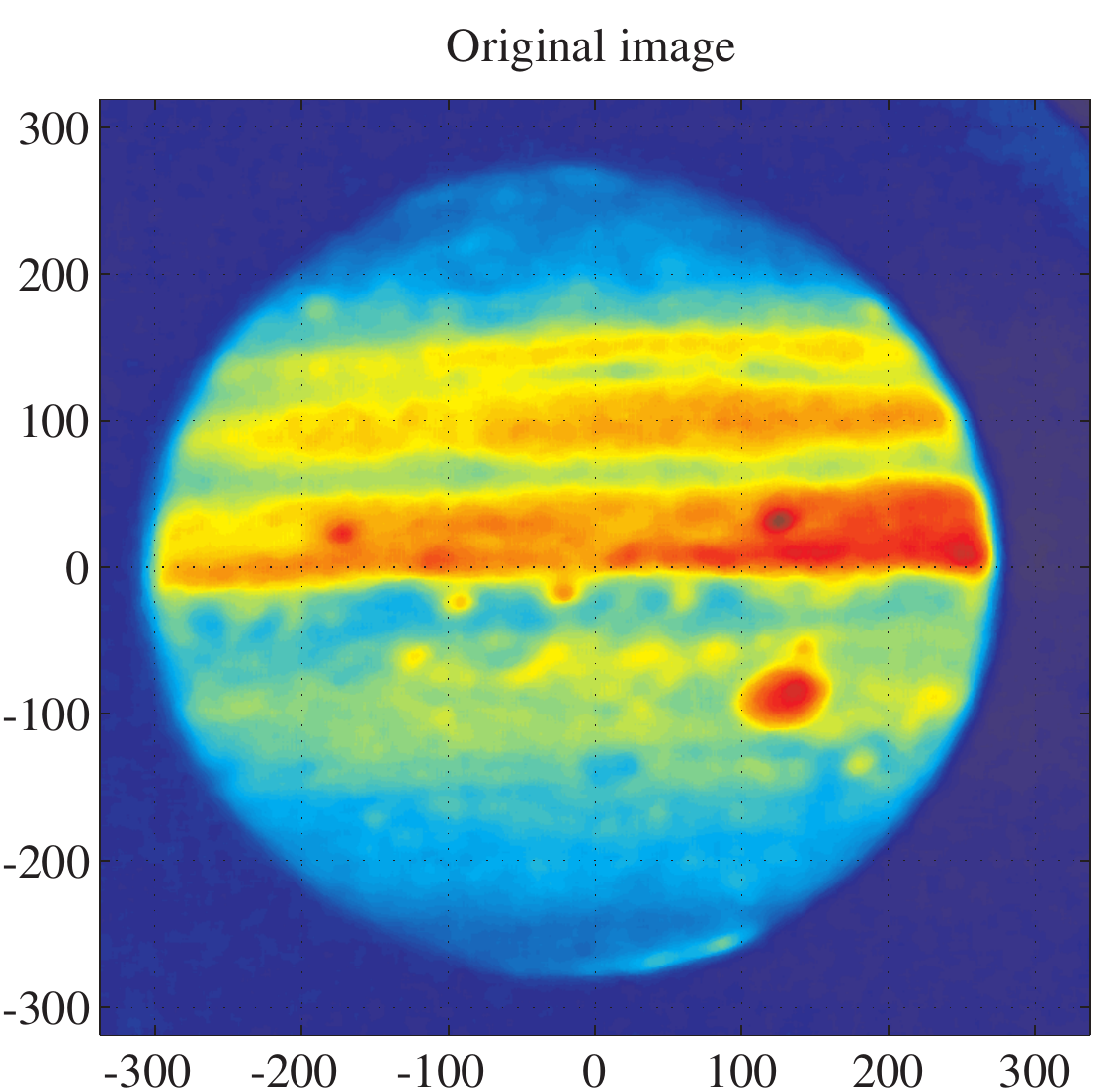}\\[0pt]
\includegraphics[width=0.99\figwidth]{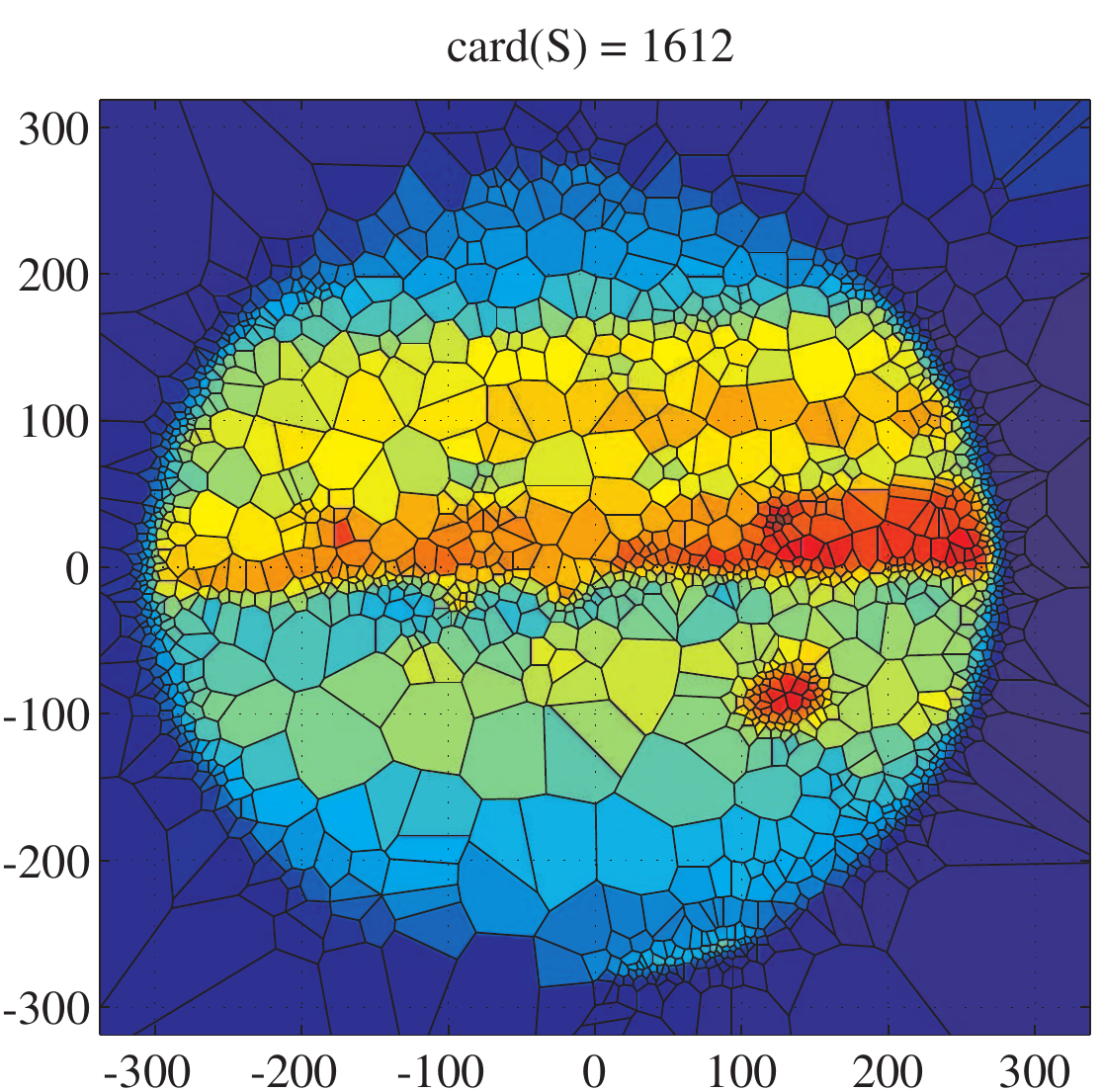}
\caption{In the upper panel, median-filtered \IR image collected by \UKIRT
(in arbitrary intensity unit) and in the lower panel the resulting Voronoi
tessellation from the \VOISE algorithm. 
The \unit[1612]{tiles} of the tessellation are uniformly coloured using
the median intensity of the pixels that are lying within each polygon.
The axes are labelled in pixels unit and the point with coordinates
$(0,0)$ is the centre of the image. The colour code for both images is
shown in \Fig{fig:fit1}. }
\label{fig:voise1}
\end{figure}

The image presented here (see upper panel of \Fig{fig:voise1}) to illustrate
the location method of a complete
planetary disc has been obtained using the \unit[3.8]{m} \UKIRT at Mauna Kea
observatory, Hawaii, with the near-\IR \UIST guide camera \citep{ramsay:2004}.
This image has not been flux calibrated but the sky background noise
has been subtracted, and the intensities are thus in arbitrary units.

\UIST is a \unit[1\mbox{--}5]{\mu m}
imager-spectrometer with a \unit[1024{\times}1024]{pixels} \chem{InSb}
array.
In imaging mode there are two plate scales available, 
with resolution \unit[0.12]{\arcsec/pixel} and 
\unit[0.06]{\arcsec\;pixel^{-1}}, giving fields of view of
\unit[2{\times}2]{\arcmin^2} and \unit[1{\times}1]{\arcmin^2} respectively.

\UIST was used to observe Jupiter at a resolution of 
\unit[0.12]{\arcsec\;pixel^{-1}}, with 
the Brackett alpha filter (\unit[50]{per cent} cut-on at \unit[4.024]{\mu m}
and \unit[50]{per cent} cut-off at \unit[4.078]{\mu m}) in exposures of
\unit[10]{s}. The Brackett line is an \IR emission line of the \chem{H}
atom. Thus this emission should contribute many of the photons as well as
\chem{H_3^+}.
In this part of the \IR spectrum,
the emission of the giant planets is dominated by several lines of
\chem{H_3^+}, and the spectral measurement of individual lines allows
determination of \chem{H_3^+} temperatures and column densities of the
planet \citep{miller:2006}.
The \UIST camera was used in conjunction with the dual-beam polarimeter
module IRPOL2 for spectropolarimetry measurements under an observation
campaign of Jupiter on August 4, 2008.

\Fig{fig:voise1} illustrates phase (i) of detection of the limb
using \VOISE on an image collected by \UKIRT at \unit[10{:}13{:}00]{UT}.
The size of the image is \unit[679{\times}639]{pixels} and 
it has been pre-processed by a nonlinear filter ---a median filter--- of size
\unit[11{\times}11]{pixels} in order to lower noise in the image
\citep{gonzalez:2007}. Whenever such noise filter is used as pre-processing
to the \VOISE segmentation, the size of the mask should be chosen to be larger 
than the minimum seed distance $d_m$ to be of any effect.
The main idea of this filter is to slide a window with specified size
and replace each centre pixel of the
window by the median of the pixels lying in the window.
The \VOISE parameters \citep{guio:2009a} are
(i) division phase: $d^2_m=\unit[9]{pixels^2}$, 
$p_\mathrm{D}=\unit[97]{per cent}$ (ii) merging phase:
$p_\mathrm{M}=\unit[50]{per cent}$, $\Delta\mu=\unit[20]{per cent}$ and
$\Delta\mathcal{H}=\unit[30]{per cent}$ (iii) two iterations in the
regularisation phase. The resulting segmentation contains
\unit[1612]{polygons} (lower panel in \Fig{fig:voise1}). Note the compactness of the polygons in regions with
small length scales along the limb and near  the equator.

\begin{figure}
\includegraphics[width=0.99\figwidth]{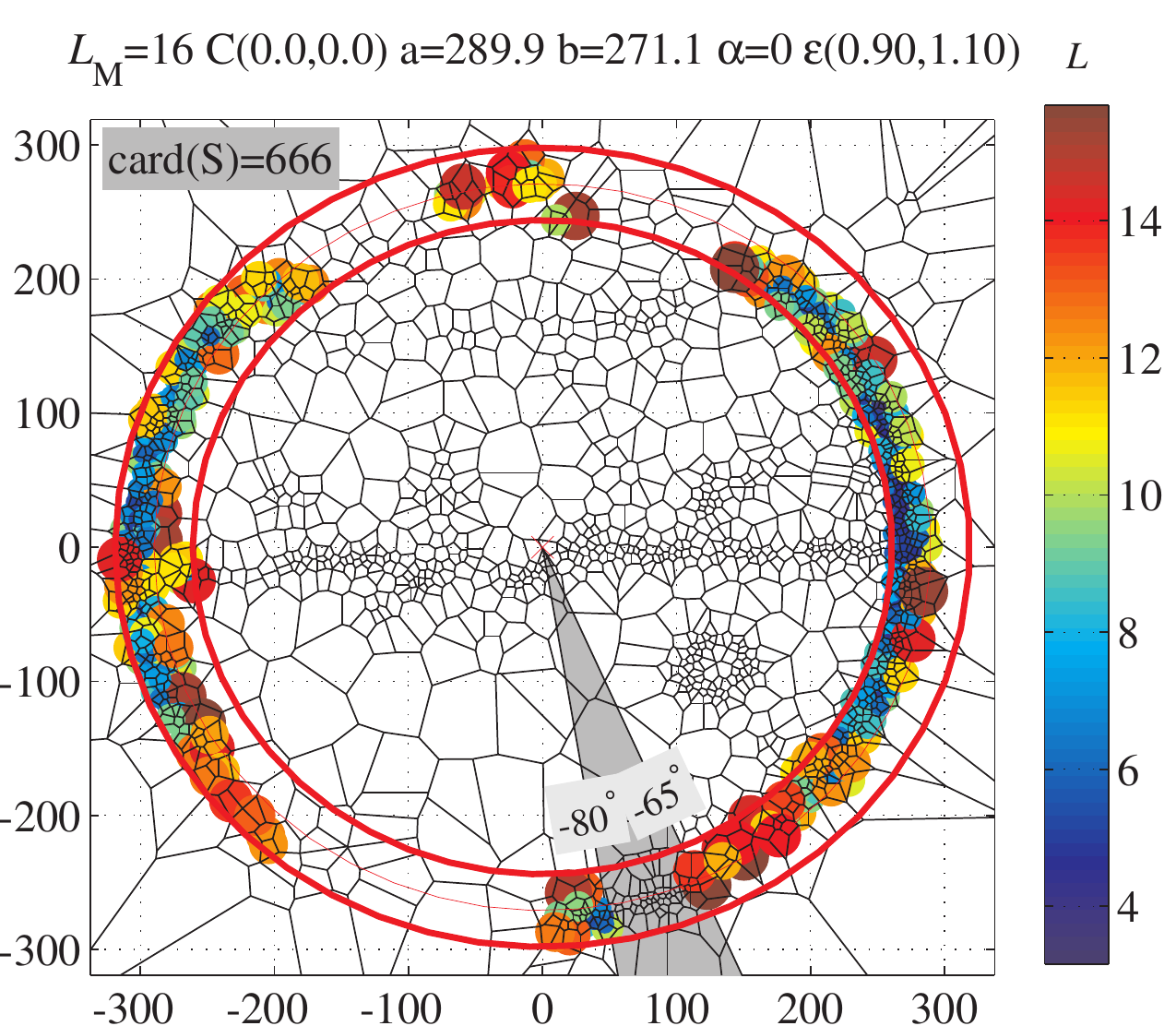}\\[0pt]
\includegraphics[width=0.99\figwidth]{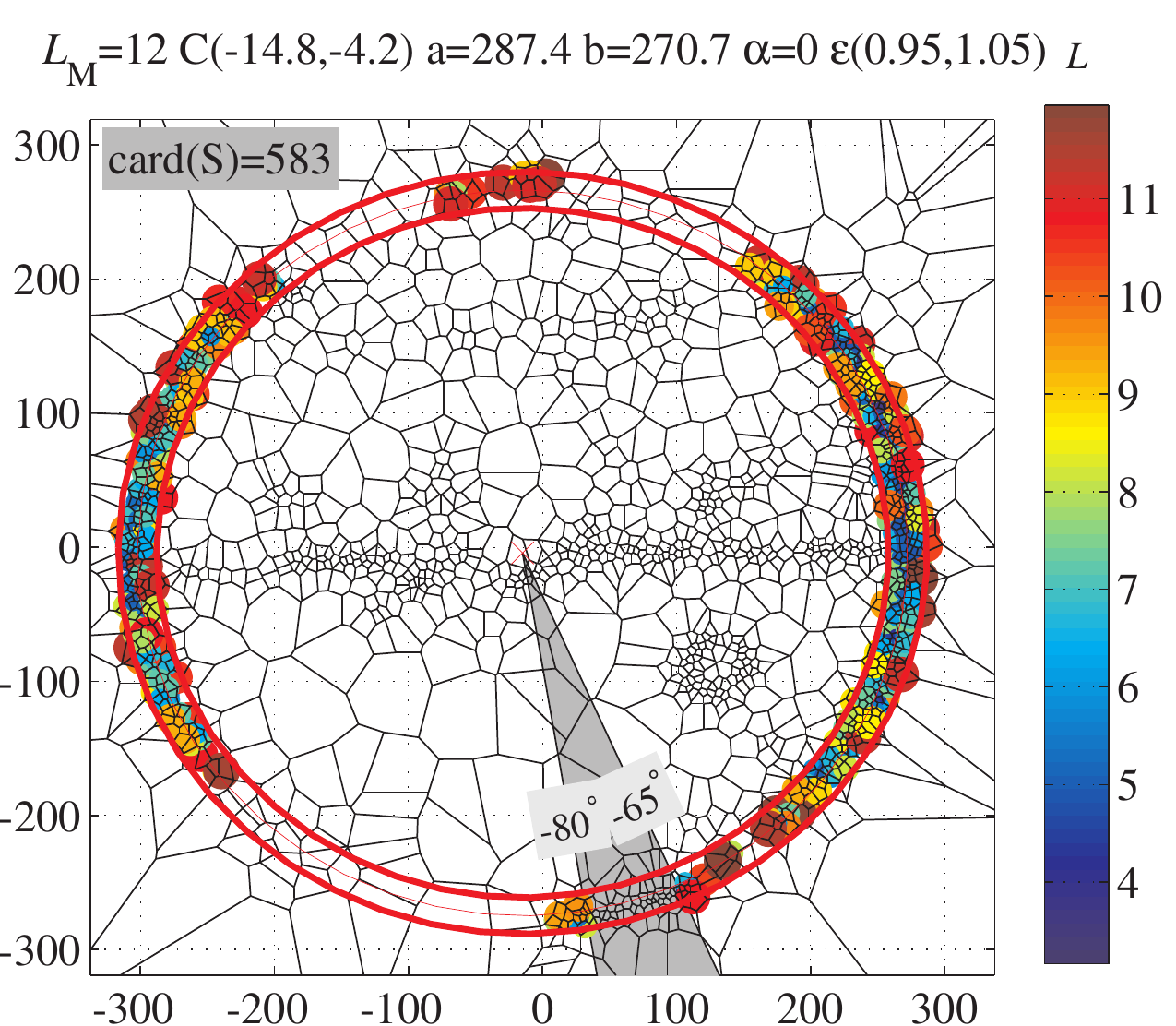}
\caption{Selection of the seeds from the \VOISE tessellation for two
iterations of the phases (ii) and (iii). During the first iteration
(upper plot) $\unit[666]{seeds}$ are selected as neighbours of the
limb while for the second iteration the numbers of seeds considered for the
fit is reduced to \unit[583]{seeds}. The size of each coloured marker is
proportional
to the surface area of the selected polygon. The limits of the torus are shown
in thick red lines while the nominal ellipse is shown as a thin red line and
the red cross is the centre of the torus.}
\label{fig:select1}
\end{figure}

\begin{table*}
\caption{Parameters of the fitted ellipse resulting from two iterations of
phases (ii) and (iii). The ellipses are shown in \Fig{fig:fit1}.
Note that the tilt angle has not been fitted and has been fixed to $\alpha=0$.
\# iter is the number of iterations performed in order to converge with 
the prescribed tolerance and the normalised $\chi^2$ provides an
indication of the goodness of the fit. The ``guess'', ``fit~1'' and
``fit~2'' ellipses are shown in \Fig{fig:fit1}.}
\label{paramfit1}
\centerline{
\begin{tabular}[b]{lccccccc}
\hline\hline 
& $\xc$ & $\yc$ & $a$ & $b$ & $\alpha$ & \# iter & $\chi^2$ \\
\hline 
\csname @@input\endcsname alljup_UT1013/fitres.dat \\
\hline\hline
\end{tabular}}
\end{table*}

\begin{figure}
\includegraphics[width=0.99\figwidth]{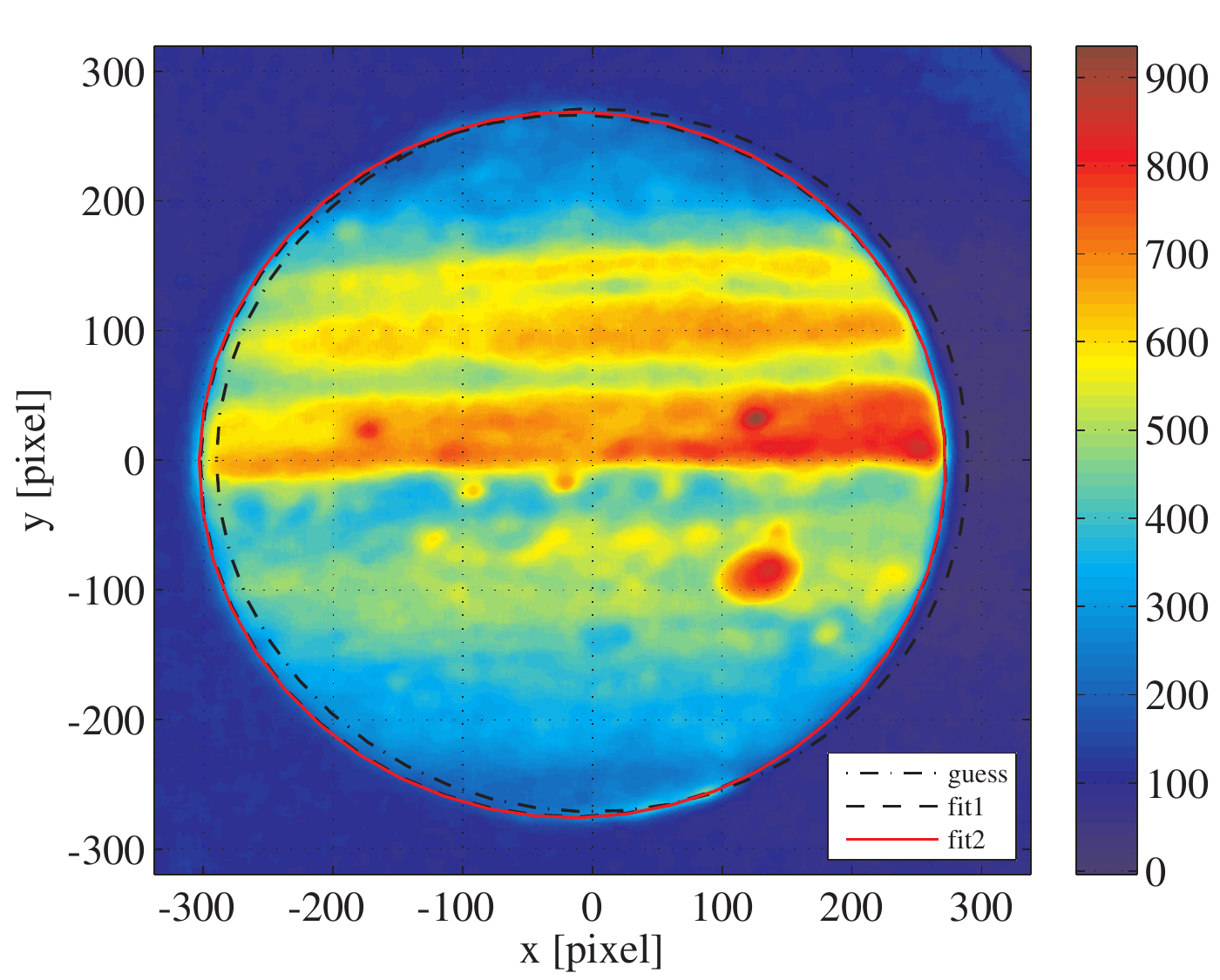}
\caption{Initial guess for the ellipse and the ellipses resulting from the two
successive fits for the original image shown in \Fig{fig:voise1}. The points
selected for each fit are as shown in \Fig{fig:select1}.}
\label{fig:fit1}
\end{figure}

\Fig{fig:select1} illustrates phase (ii) of selecting the set of
points from the segmentation to be used as the neighbourhood of the planetary
limb. The upper panel shows the first iteration, using crude
estimates of the ellipse parameters to define the large torus with
$\varepsilon_m=0.9$ and $\varepsilon_M=1.1$, and a relatively large scale
length parameter $\mathcal{L}_M=\unit[16]{pixels}$. 
The lower panel shows the selection process for the second iteration with
parameters provided by the result of the first fit. The torus has been
re-centred and its thickness reduced by setting $\varepsilon_m=0.9$
and $\varepsilon_M=1.05$. The scale length parameter has also been reduced
to $\mathcal{L}_M=\unit[12]{pixels}$ (to be compared to
$d_m=\unit[3]{pixels}$).
In both iterations, seeds in the neighbourhood of the faint emission outside
the limb near the South pole have been rejected whenever the polar angle of
the seed $\varphi$ with respect to the centre $C(\xc,\yc)$ is in the range
$\unit[-80]{^\circ}{<}\varphi{<}\unit[-65]{^\circ}$,
i.e.\ the seed lies inside the grey shaded sector depicted in
\Fig{fig:select1}.

\Fig{fig:fit1} illustrates phase (iii) consisting of the nonlinear
fitting of the selected points (shown in \Fig{fig:select1}) to an ellipse
in parametric form given by \Eq{eq:ellipse}.
The planetary disc modelled as a single ellipse is justified in the situation
where the disc is nearly fully illuminated, as is the case here. 
The curve labelled ``guess'' corresponds to crude estimates of the ellipse
parameters, i.e.\ the coordinates of the planet centre correspond to the
centre of the image and the equatorial and polar radii are derived using
\SPICE. The curve
``fit1'' is the curve with parameters after first fit, i.e.\ with the seeds
as seen in the upper panel of \Fig{fig:select1} and the the curve labelled
``fit2'' corresponds to seeds as seen in the lower panel of
\Fig{fig:select1}. 

The parameters, error estimates and fitting parameters
are given in \Table{paramfit1}.
It is interesting to note that the estimated parameters related to the
$x$-direction ($\xc$ and $a$) have smaller errors compared to the parameters
relates to the $y$-direction ($\yc$ and $b$) which is a consequence of the
large sampling of seeds around the equator.
We have also checked, for consistency, that the curve parameters $\{\phi_i\}$
and the global parameters $\vec{p}$ together with their errors
$\{\Delta\phi_i\}$ and $\Delta\vec{p}$ provide error in positioning of the
points consistent with the values used for the weights $\sigma_i$ in
\Eq{eq:minfun}. 
\comment{thus confirming that the length scale derived from the
surface area of the polygons provides good estimates of error in location.}

\begin{figure}
\includegraphics[width=0.99\figwidth]{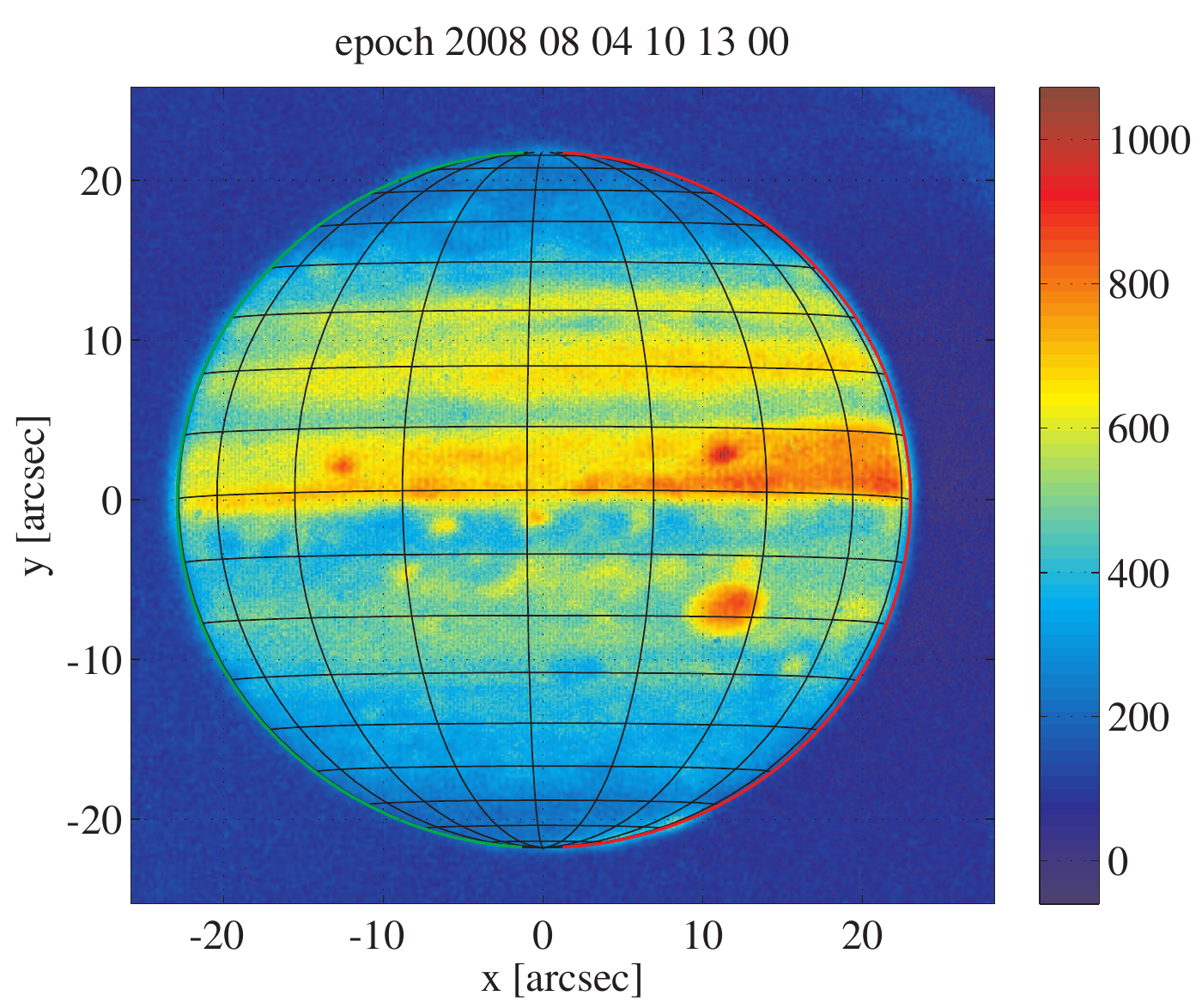}
\caption{Image scaled in \unit{\arcsec} and the projection in the sky-plane of
a latitude-longitude grid of Jupiter computed using data
from \SPICE and the fitted parameters of Jupiter's disc.}
\label{fig:limb1}
\end{figure}

\Fig{fig:limb1}  presents the image with a (planetocentric) latitude-longitude
grid (with \unit[10]{^\circ} step in latitude and \unit[20]{^\circ} in
longitude)
computed using the coordinates of the centre of Jupiter's disc obtained
from the nonlinear fitting and the projection geometry computed using 
\SPICE. The \CML of Jupiter at the time of the observation is
\unit[157.5]{deg}. The limb is shown on the right side of the
planet (red solid line) while the terminator is shown on the left side
(green solid line).
The parameters are $\latobs=\unit[-1.5]{^\circ}$,
$\latsun=\unit[-1.4]{^\circ}$ and $\dlon=\unit[5.3]{^\circ}$. The equatorial
radius for the final fit is $\re=\unit[70958]{km}$ with an eccentricity 
$e=0.32$ while the values provided by \SPICE are $\re=\unit[71492]{km}$ and
$e=0.35$. The illuminated limb in the considered near-\IR waveband is
thus slightly smaller than Jupiter's \unit[1]{Bar} pressure surface as given
by \SPICE. 

\subsection{Circle for partial planet}

\begin{figure}
\includegraphics[width=0.99\figwidth]{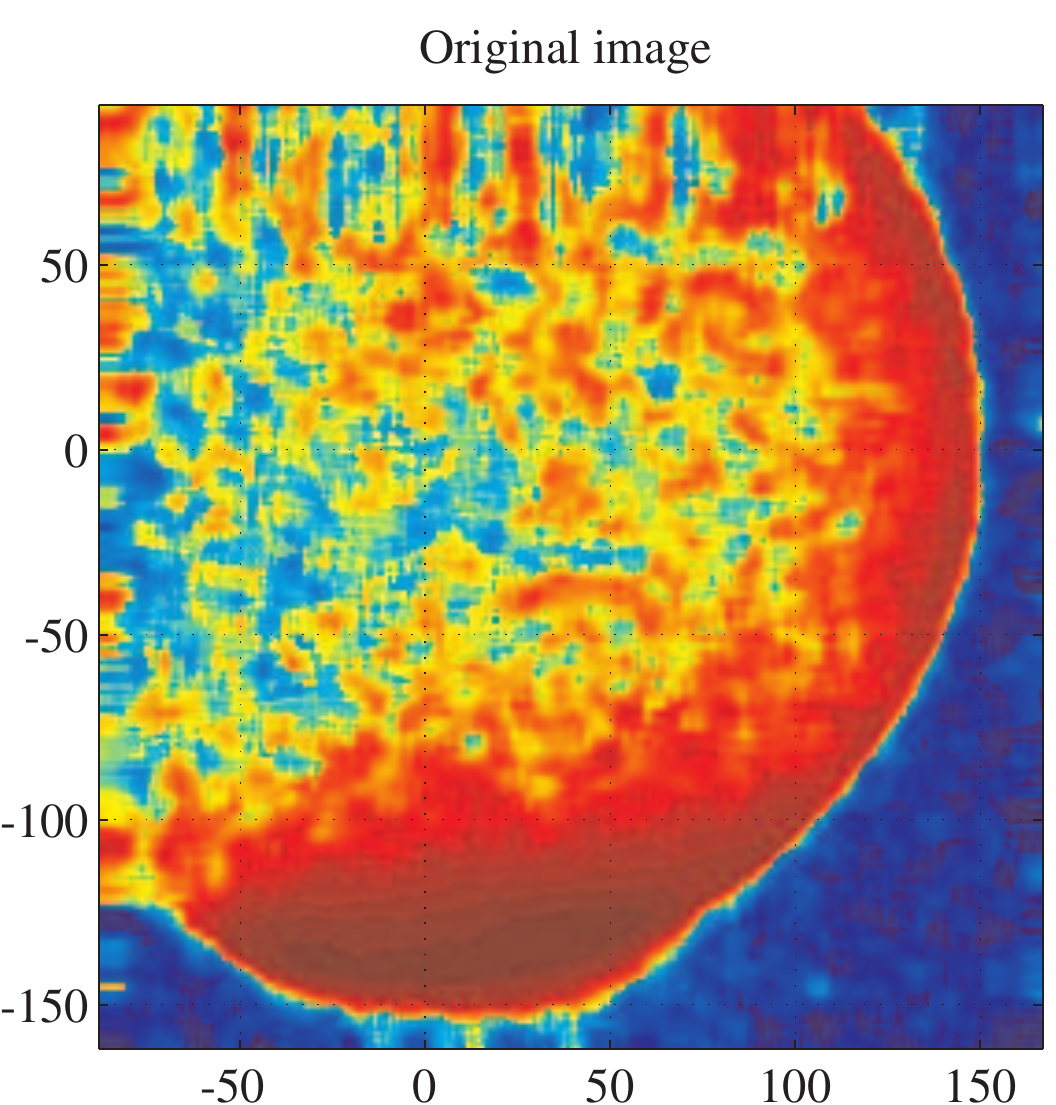}\\[0pt]
\includegraphics[width=0.99\figwidth]{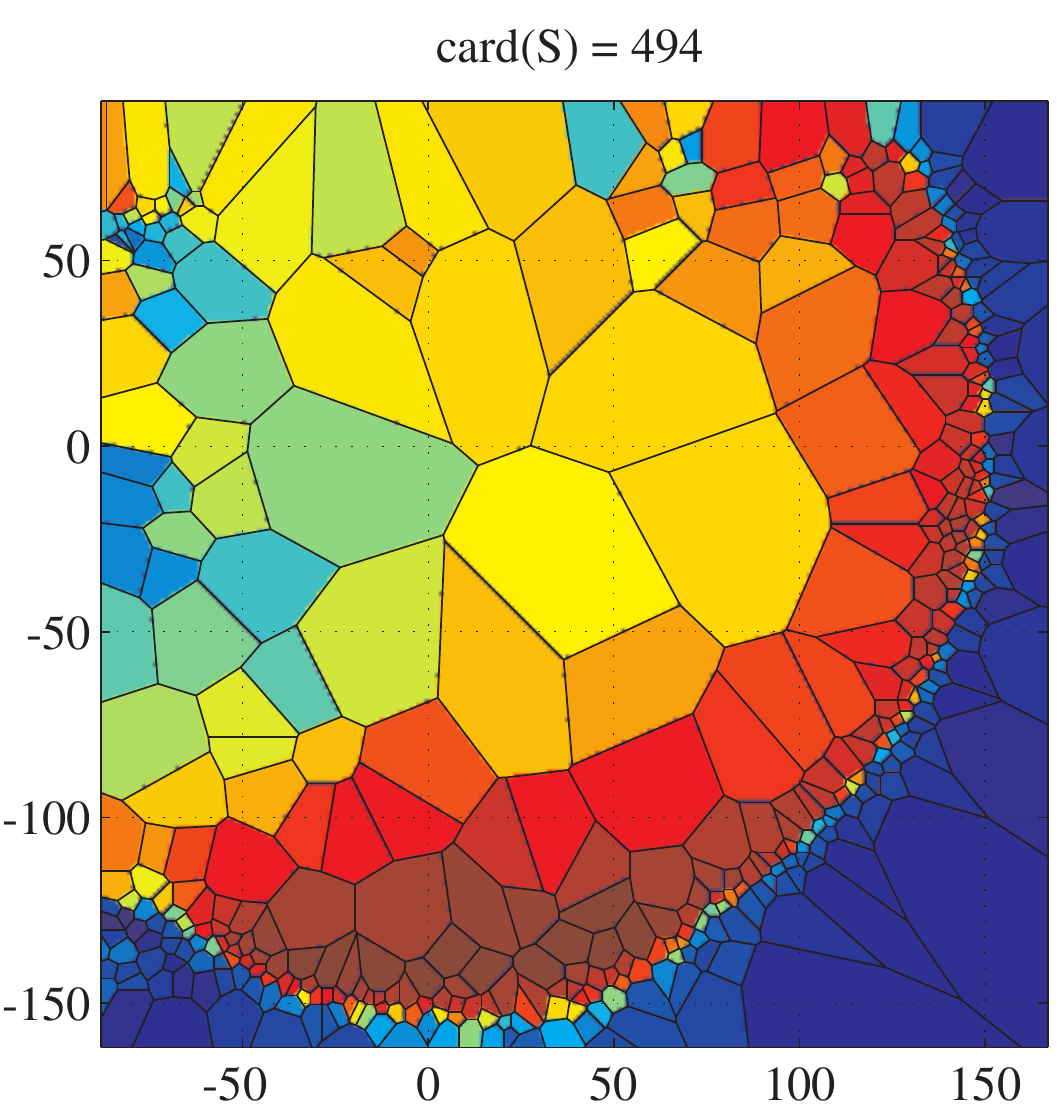}
\caption[]{In the top panel, \IRTF image processed by a median filter
followed by histogram equalisation (in arbitrary intensity unit). In the
lower panel the result of the segmentation by \VOISE. The colour code for
the \unit[494]{tiles} is the same as for the segmentation presented in the 
lower panel in \Fig{fig:voise1}.
The axes are labelled in pixels unit and coordinates $(0,0)$ correspond to
the centre of the planet provided by the original \IRAF data reduction,
taking into account an estimate based on the telescope pointing
\citep{satoh:1999}. The colour code for both images is
shown in \Fig{fig:fit2}. }
\label{fig:voise2}
\end{figure}

The image presented in this section (upper panel in \Fig{fig:voise2}) 
has been chosen to illustrate the 
case with partial occlusion of the planetary disc. It was collected with
NASA's \unit[3.8]{m} \IRTF at Mauna Kea
observatory, Hawaii using the \NSFCAM imaging facility \citep{shure:1994} 
at wavelength \unit[3.43]{\mu m} (a wavelength sensitive to \chem{H_3+}). 
The image was collected during a campaign on June 28, 1995 at
\unit[11{:}14{:}52]{UT}.
This image has not been flux calibrated but the sky background noise
has been subtracted, and the intensities are thus in arbitrary units.

The NSFCAM is a \unit[1\hbox{--}5]{\mu m} imager with a 
\unit[256{\times}256]{pixels} \chem{InSb} detector.
Three different magnifications are available: \unit[0.3]{\arcsec\;pixel^{-1}},
\unit[0.15]{\arcsec\;pixel^{-1}} and \unit[0.06]{\arcsec\;pixel^{-1}} 
corresponding to 
a field of view of \unit[76.8]{\arcsec}, \unit[37.9]{\arcsec} and
\unit[14.1]{\arcsec} respectively.
The \NSFCAM has been upgraded (\NSFCAMTWO) with a 
\unit[2048{\times}2048]{pixels} Hawaii-2RG detector. The image scale will be
\unit[0.04]{\arcsec\;pixel^{-1}} with field of view 
\unit[80{\times}80]{\arcsec^2}.

\begin{figure}
\includegraphics[width=0.99\figwidth]{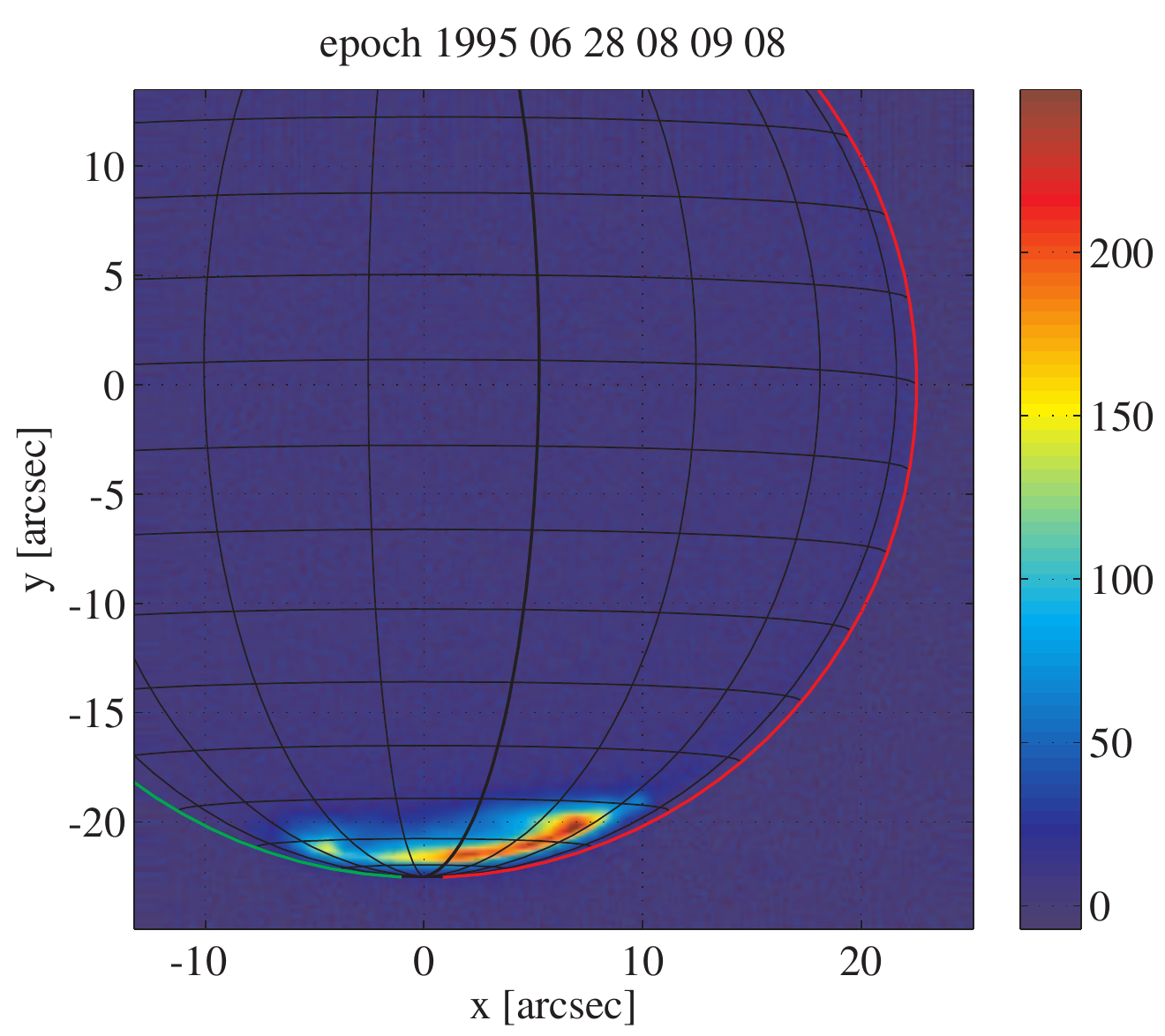}
\caption{Original image scaled in \unit{\arcsec} together with 
the sky-plane projection of a latitude-longitude grid
computed using data
from \SPICE and the fitted parameters of Jupiter's disc, i.e.\ the 
circle parameters.}
\label{fig:limb2}
\end{figure}

\Fig{fig:voise2} shows the results of phase (i) of the method
using \VOISE on an image collected by \UKIRT at \unit[0721]{UT}.
The image has size \unit[256{\times}256]{pixels}.
Note that the image has been pre-processed by a median filter of size
\unit[7{\times}7]{pixels} to lower noise level, followed by a
histogram equalisation \citep{gonzalez:2007}. The histogram equalisation
is performed in order to increase the global contrast of the original image 
(which is shown in \Fig{fig:limb2}).
It consists of a nonlinear adjustment of the intensities in order to better
distribute the image intensity histogram and
accomplishes this by effectively ``spreading
out'' the most frequent intensity values.
Alternatively, the contrast in the low intensity range can also be enhanced
by taking the logarithm of the ratio of the image pixels relative to the
estimated  noise level, if available. Note that we haven't pre-processed the
\UKIRT image for the first example as the limb boundary was already
substantially more intense than the background.
The \VOISE parameters have been set to (i) division phase:
$d^2_m=\unit[4]{pixels^2}$,
$p_\mathrm{D}=\unit[98]{per cent}$ (ii) merging phase:
$p_\mathrm{M}=\unit[50]{per cent}$, $\Delta\mu=\unit[20]{per cent}$ and
$\Delta\mathcal{H}=\unit[30]{per cent}$ (iii) two iterations  in the
regularisation phase.

\begin{figure}
\includegraphics[width=0.99\figwidth]{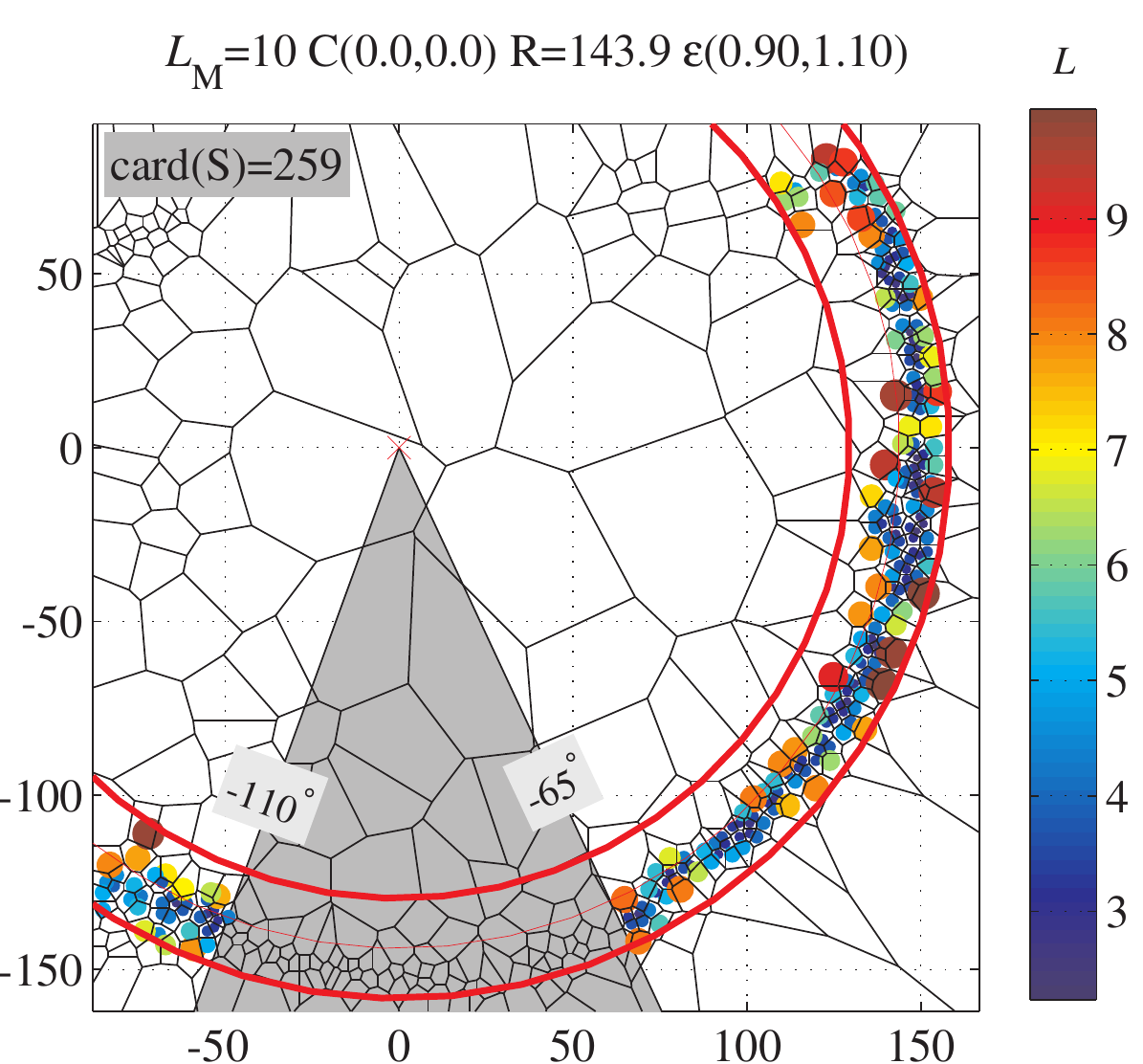}\\[0pt]
\includegraphics[width=0.99\figwidth]{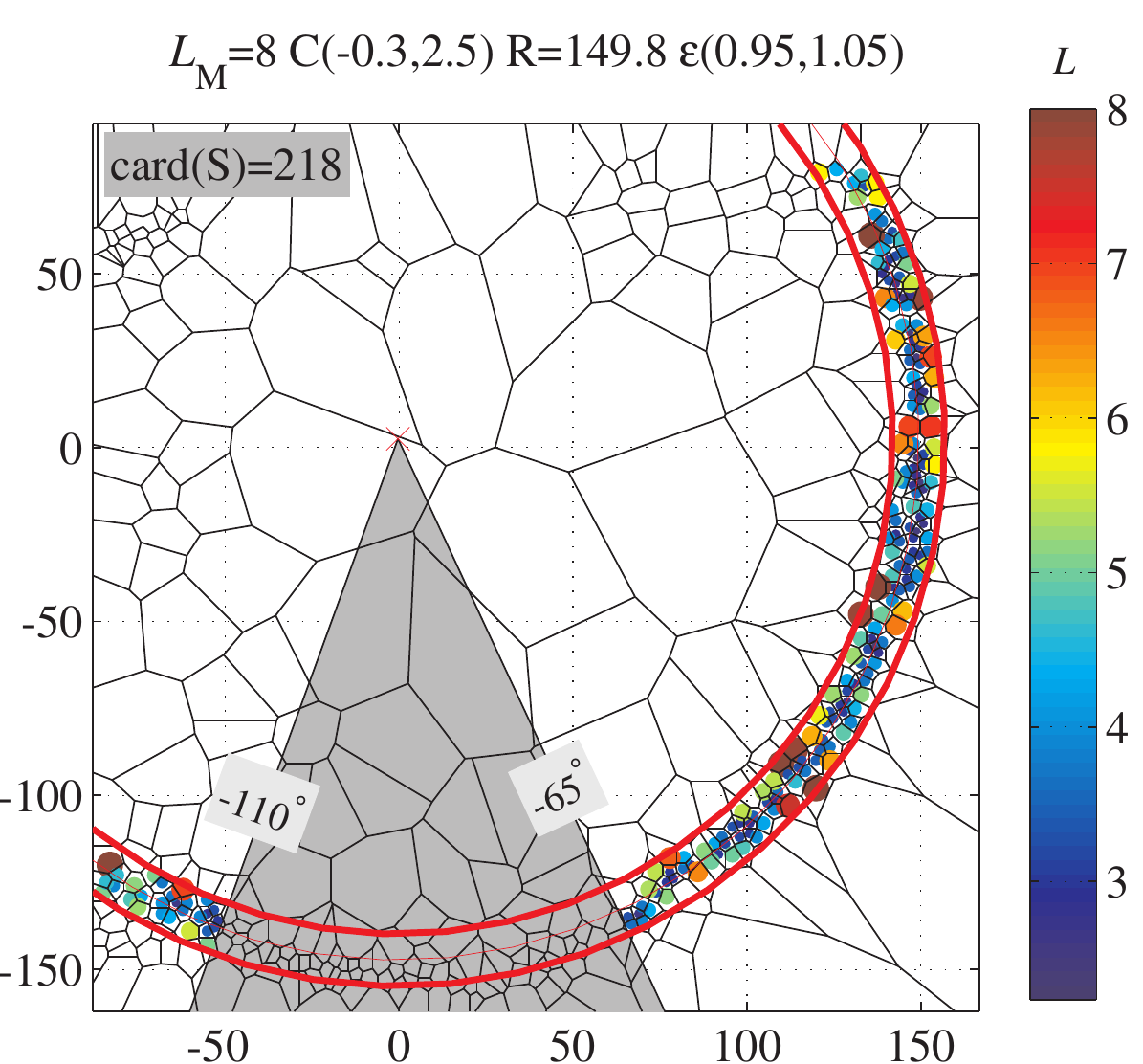}
\caption{Seeds selection for two iterations of phase (ii). For details
see \Fig{fig:select1}.}
\label{fig:select2}
\end{figure}

\Fig{fig:select2} illustrates phase (ii) of selecting the set of
points from the Voronoi map to be used as within the limb neighbourhood.
Note that the points with polar angle $\varphi$ with respect to the centre
estimate $C(\xc,\yc)$ such that
$\unit[-115]{^\circ}{<}\varphi{<}\unit[-65]{^\circ}$ (inside the grey
shaded sector in \Fig{fig:select2}) are
filtered out to avoid bias from the seeds corresponding to the the emission
outside the limb which has been highlighted by the histogram equalisation.

\begin{figure}
\includegraphics[width=0.99\figwidth]{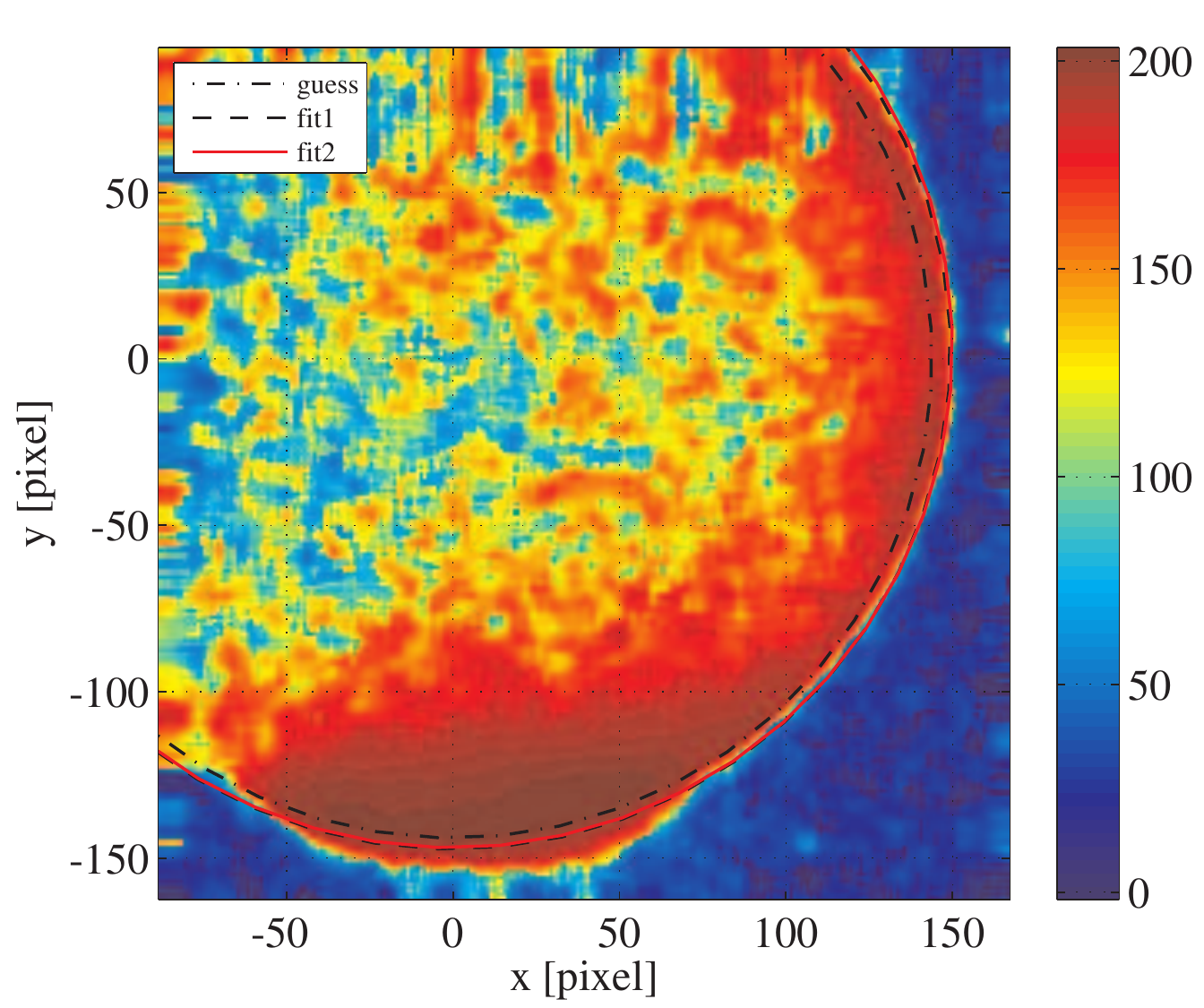}
\caption{Initial guess for the circle and the circles ``fit~1'' and
``fit~2'' resulting from two iterations of the phases (ii) and (iii). 
The selected points for each iteration are shown in \Fig{fig:select2}.}
\label{fig:fit2}
\end{figure}

\Fig{fig:fit2} illustrates phase (iii) consisting of the nonlinear
fitting of the selected points (shown in \Fig{fig:select2}) to a circle
in parametric form \Eq{eq:circle}. The global parameters of the circle
fitted together with estimates for the error are given in \Table{paramfit2}.
The circle as a model for the disc is justified in situations where only
a portion of the planetary disc is in the field of view, as it is the case in
the present image.
Note also that in this case the distribution of the seeds is uniform 
from the equator to the South pole and therefore the estimated parameters
related to the $x$-direction have similar errors as those 
related to the $y$-direction.

\begin{table}
\caption{Resulting parameters for the circle during the two iterations of
fit. The number of iterations and the resulting normalised $\chi^2$ are also
given. The corresponding ``guess'', ``fit~1'' and ``fit~2'' circles are
shown in \Fig{fig:fit2}.}
\label{paramfit2}
\centerline{
\begin{tabular}[b]{lccccc}
\hline\hline
& $\xc$ & $\yc$ & $R$ & \# iter & $\chi^2$ \\
\hline
\csname @@input\endcsname 28jn014s/fitres.dat \\
\hline\hline
\end{tabular}
}
\end{table}

\Fig{fig:limb2} presents the image with the same (planetocentric) 
latitude-longitude grid resolution as in \Fig{fig:limb1}
calculated with the coordinates of the centre and radius of Jupiter's disc
obtained from the nonlinear fitting, and the projection geometry from
\SPICE. The \CML of Jupiter at the time of the observation is
\unit[13.5]{deg}. 
The limb and terminator are shown as red and green solid lines respectively,
and the thick black line is the noon-meridian.
The parameters provided by SPICE are $\latobs=\unit[-2.9]{^\circ}$,
$\latsun=\unit[-2.8]{^\circ}$ and $\dlon=\unit[5.3]{^\circ}$. 
The radius for the final fit (``fit~2'') is $\re=\unit[72466]{km}$. 
The illuminated limb in this waveband is slightly larger than 
Jupiter's \unit[1]{bar} pressure surface.
We also tried to fit an ellipse and it leads to a similar estimate for the
centre coordinates but with larger error bars due to the increased degrees
of freedom. 
\comment{
[NOTE: in this case the fitted equatorial radius is
$\re=\unit[71923]{km}$ and the polar radius is larger than the equatorial 
radius $\rp=\unit[73017]{km}$! This means an eccentricity in the $y_s$-axis 
$\sim0.17$. Should we comment that as \chem{H_3^+} polar emission
``dilating'' the pole???]
}

\section{Discussion}
\label{sec:discussion}

We have presented a novel semi-automatic method to estimate accurately, 
and objectively, the disc parameters in an image of an illuminated planetary
disc. The method is based on the ``best'' fit of a set of points selected
from a segmentation map generated by \VOISE to a curve described in a
parametric form. 

The segmentation phase can be improved by pre-processing the image 
using different techniques such as noise filtering and
contrast adjustment.

Basic shapes to describe the boundary of a planetary disc include the
circle and the ellipse. 
We also provide analytic expressions for the projection in the sky-plane of the
limb and terminator of a planet modelled as ellipsoid.
These expressions can easily be used to describe both limb and terminator as
single curve in parametric form.

Note that the \VOISE algorithm generates ``intermediate'' tessellations,
one at the end of the division phase and one at the end of the merging
phase. It is worth noting that fitting an ellipse gives the best result
(smallest $\chi^2$) for the regularised tessellation (i.e.\ after merging),
but errors in the
fitted parameters are smaller
when considering the map at the end of the division phase. The largest
$\chi^2$ is obtained for the map generated at the end of the merging phase.
This confirms that 
the  tessellation obtained after division is optimum for our purposes.
The reason for this is that \VOISE merging generates more regular
polygons, but very slightly degrades the position information from the
division phase.

We have shown that our novel objective method to locate the planetary
disc on images provides improved estimates of the centre position (as compared
to the guide star catalogue) as well as the altitude when the disc is
illuminated for the corresponding observational waveband.

We also showed that the use of histogram equalisation
enhances the auroral emission outside the limb and therefore allows a
better and unbiased estimate of the limb by allowing removal of 
points from this auroral emission region.

The software implementing this method is written in \matlab and can be made
available by request to the authors.

\section*{Acknowledgements}

We would like to thank M. Lystrup who kindly provided the \UKIRT images.

The United Kingdom Infrared Telescope is operated by the Joint Astronomy
Centre on behalf of the Science and Technology Facilities Council of the
U.K.

We also like to thank J.E.P. Connerney and T. Satoh for making the
\IRTF data available.

This work uses data acquired at the NASA \IRTF,
which is operated by the University of Hawaii under
Cooperative Agreement no.\ NNX-08AE38A with the National Aeronautics and
Space Administration, Science Mission Directorate, Planetary Astronomy
Program. 

\bibliography{abbrevs,research,books}

\begin{thebibliography}{29}
\expandafter\ifx\csname natexlab\endcsname\relax\def\natexlab#1{#1}\fi

\bibitem[{{Acton}(1996)}]{acton:1996}
{Acton} C.~H., 1996, Planet. Space Sci., 44, 65

\bibitem[{{Badman} {et~al.}(2008){Badman}, {Cowley}, {Lamy}, {Cecconi}, \&
  {Zarka}}]{badman:2008}
{Badman} S.~V., {Cowley} S.~W.~H., {Lamy} L., {Cecconi} B., {Zarka} P., 2008,
  Ann. Geophysic\ae, 26, 3641

\bibitem[{Bard(1974)}]{bard:1974}
Bard Y., 1974, Non linear parameter estimation. Academic Press, New York, iSBN
  0-12-078250-2

\bibitem[{{Bonfond} {et~al.}(2007){Bonfond}, {G{\'e}rard}, {Grodent}, \&
  {Saur}}]{bonfond:2007}
{Bonfond} B., {G{\'e}rard} J.-C., {Grodent} D., {Saur} J., 2007, Geophys. Res.
  Lett., 34, 6201

\bibitem[{{Bonfond} {et~al.}(2009){Bonfond}, {Grodent}, {G{\'e}rard},
  {Radioti}, {Dols}, {Delamere}, \& {Clarke}}]{bonfond:2009}
{Bonfond} B., {Grodent} D., {G{\'e}rard} J., {Radioti} A., {Dols} V.,
  {Delamere} P.~A., {Clarke} J.~T., 2009, J. Geophys. Res., 114, 7224

\bibitem[{Bookstein(1979)}]{bookstein:1979}
Bookstein F.~L., 1979, Comput. Graph. Image Process., 9, 56

\bibitem[{{Bunce} {et~al.}(2008){Bunce}, {Arridge}, {Clarke}, {Coates},
  {Cowley}, {Dougherty}, {G{\'e}rard}, {Grodent}, {Hansen}, {Nichols},
  {Southwood}, \& {Talboys}}]{bunce:2008}
{Bunce} E.~J., {Arridge} C.~S., {Clarke} J.~T., {Coates} A.~J., {Cowley}
  S.~W.~H., {Dougherty} M.~K., {G{\'e}rard} J., {Grodent} D., {Hansen} K.~C.,
  {Nichols} J.~D., {Southwood} D.~J., {Talboys} D.~L., 2008, J. Geophys. Res.,
  113, 9209

\bibitem[{{Clarke} {et~al.}(2002){Clarke}, {Ajello}, {Ballester}, {Ben Jaffel},
  {Connerney}, {G{\'e}rard}, {Gladstone}, {Grodent}, {Pryor}, {Trauger}, \&
  {Waite}}]{clarke:2002}
{Clarke} J.~T., {Ajello} J., {Ballester} G., {Ben Jaffel} L., {Connerney} J.,
  {G{\'e}rard} J.-C., {Gladstone} G.~R., {Grodent} D., {Pryor} W., {Trauger}
  J., {Waite} J.~H., 2002, Nature, 415, 997

\bibitem[{{Clarke} {et~al.}(2005){Clarke}, {G{\'e}rard}, {Grodent},
  {Wannawichian}, {Gustin}, {Connerney}, {Crary}, {Dougherty}, {Kurth},
  {Cowley}, {Bunce}, {Hill}, \& {Kim}}]{clarke:2005}
{Clarke} J.~T., {G{\'e}rard} J., {Grodent} D., {Wannawichian} S., {Gustin} J.,
  {Connerney} J., {Crary} F., {Dougherty} M., {Kurth} W., {Cowley} S.~W.~H.,
  {Bunce} E.~J., {Hill} T., {Kim} J., 2005, Nature, 433, 717

\bibitem[{{Dougherty} {et~al.}(1998){Dougherty}, {Dunlop}, {Prange}, \&
  {Rego}}]{dougherty:1998}
{Dougherty} M.~K., {Dunlop} M.~W., {Prange} R., {Rego} D., 1998, Planet. Space
  Sci., 46, 531

\bibitem[{Fitzgibbon {et~al.}(1999)Fitzgibbon, Pilu, \&
  Fisher}]{fitzgibbon:1999}
Fitzgibbon A., Pilu M., Fisher R., 1999, IEEE Trans. Pattern Anal. Mach.
  Intell., 21, 476

\bibitem[{Gander {et~al.}(1994)Gander, Golub, \& Strebel}]{gander:1994}
Gander W., Golub G.~H., Strebel R., 1994, BIT Num. Math., 34, 558

\bibitem[{{Gonzalez} \& {Woods}(2007)}]{gonzalez:2007}
{Gonzalez} R.~C., {Woods} R.~E., 2007, {Digital image processing}, 3rd edn.
  Prenctice Hall, Upper Saddle River, NJ, iSBN 0240515749

\bibitem[{{Grodent} {et~al.}(2003{\natexlab{a}}){Grodent}, {Clarke}, {Kim},
  {Waite}, \& {Cowley}}]{grodent:2003b}
{Grodent} D., {Clarke} J.~T., {Kim} J., {Waite} J.~H., {Cowley} S.~W.~H.,
  2003{\natexlab{a}}, J. Geophys. Res., 108, 1389

\bibitem[{{Grodent} {et~al.}(2003{\natexlab{b}}){Grodent}, {Clarke}, {Waite},
  {Cowley}, {G{\'e}rard}, \& {Kim}}]{grodent:2003a}
{Grodent} D., {Clarke} J.~T., {Waite} J.~H., {Cowley} S.~W.~H., {G{\'e}rard}
  J.-C., {Kim} J., 2003{\natexlab{b}}, J. Geophys. Res., 108, 1366

\bibitem[{{Guio, P. and Achilleos, N.}(2009)}]{guio:2009a}
{Guio, P. and Achilleos, N.}, 2009, Mon. Not. R. Astron. Soc., 1051

\bibitem[{{Lamy} {et~al.}(2009){Lamy}, {Cecconi}, {Prang{\'e}}, {Zarka},
  {Nichols}, \& {Clarke}}]{lamy:2009}
{Lamy} L., {Cecconi} B., {Prang{\'e}} R., {Zarka} P., {Nichols} J.~D., {Clarke}
  J.~T., 2009, J. Geophys. Res., 114, 10212

\bibitem[{Marquardt(1963)}]{marquardt:1963}
Marquardt D.~W., 1963, SIAM J. Appl. Math., 11, 431

\bibitem[{{Miller} {et~al.}(2006){Miller}, {Stallard}, {Smith}, {Millward},
  {Mellin}, {Lystrup}, \& {Aylward}}]{miller:2006}
{Miller} S., {Stallard} T., {Smith} C., {Millward} G., {Mellin} H., {Lystrup}
  M., {Aylward} A., 2006, Phil. Trans. Roy. Soc. London A, 364, 3121

\bibitem[{{Nichols} {et~al.}(2008){Nichols}, {Clarke}, {Cowley}, {Duval},
  {Farmer}, {G{\'e}rard}, {Grodent}, \& {Wannawichian}}]{nichols:2008}
{Nichols} J.~D., {Clarke} J.~T., {Cowley} S.~W.~H., {Duval} J., {Farmer} A.~J.,
  {G{\'e}rard} J.-C., {Grodent} D., {Wannawichian} S., 2008, J. Geophys. Res.,
  113, 11205

\bibitem[{{Prang{\'e}} {et~al.}(1998){Prang{\'e}}, {Rego}, {Pallier},
  {Connerney}, {Zarka}, \& {Queinnec}}]{prange:1998}
{Prang{\'e}} R., {Rego} D., {Pallier} L., {Connerney} J., {Zarka} P.,
  {Queinnec} J., 1998, J. Geophys. Res., 103, 20195

\bibitem[{{Prang{\'e}} {et~al.}(1996){Prang{\'e}}, {Rego}, {Southwood},
  {Zarka}, {Miller}, \& {Ip}}]{prange:1996}
{Prang{\'e}} R., {Rego} D., {Southwood} D., {Zarka} P., {Miller} S., {Ip} W.,
  1996, Nature, 379, 323

\bibitem[{{Ramsay Howat} {et~al.}(2004){Ramsay Howat}, {Todd}, {Leggett},
  {Davis}, {Strachan}, {Borrowman}, {Ellis}, {Elliot}, {Gostick}, {Kackley}, \&
  {Rippa}}]{ramsay:2004}
{Ramsay Howat} S.~K., {Todd} S., {Leggett} S., {Davis} C., {Strachan} M.,
  {Borrowman} A., {Ellis} M., {Elliot} J., {Gostick} D., {Kackley} R., {Rippa}
  M., 2004, in Presented at the Society of Photo-Optical Instrumentation
  Engineers (SPIE) Conference, Vol. 5492, Society of Photo-Optical
  Instrumentation Engineers (SPIE) Conference Series, {A.~F.~M.~Moorwood \&
  M.~Iye}, ed., pp. 1160--1171

\bibitem[{{Satoh} \& {Connerney}(1999)}]{satoh:1999}
{Satoh} T., {Connerney} J.~E.~P., 1999, Icarus, 141, 236

\bibitem[{{Shure} {et~al.}(1994){Shure}, {Toomey}, {Rayner}, {Onaka}, \&
  {Denault}}]{shure:1994}
{Shure} M.~A., {Toomey} D.~W., {Rayner} J.~T., {Onaka} P.~M., {Denault} A.~J.,
  1994, in Society of Photo-Optical Instrumentation Engineers (SPIE) Conference
  Series, Vol. 2198, Society of Photo-Optical Instrumentation Engineers (SPIE)
  Conference Series, {D.~L.~Crawford \& E.~R.~Craine}, ed., pp. 614--622

\bibitem[{{Talboys} {et~al.}(2009){Talboys}, {Arridge}, {Bunce}, {Coates},
  {Cowley}, \& {Dougherty}}]{talboys:2009}
{Talboys} D.~L., {Arridge} C.~S., {Bunce} E.~J., {Coates} A.~J., {Cowley}
  S.~W.~H., {Dougherty} M.~K., 2009, J. Geophys. Res., 114, 6220

\bibitem[{Taubin(1991)}]{taubin:1991}
Taubin G., 1991, IEEE Trans. Pattern Anal. Mach. Intell., 13, 1115

\bibitem[{{Wannawichian} {et~al.}(2008){Wannawichian}, {Clarke}, \&
  {Pontius}}]{wannawichian:2008}
{Wannawichian} S., {Clarke} J.~T., {Pontius} D.~H., 2008, J. Geophys. Res.,
  113, 7217

\bibitem[{Yuen {et~al.}(1989)Yuen, Illingworth, \& Kittler}]{yuen:1989}
Yuen H.~K., Illingworth J., Kittler J., 1989, Image Vis. Comput., 7, 31

\end{thebibliography}

\bsp

\label{lastpage}

\end{document}